\documentclass[prb,superscriptaddress,longbibliography,twocolumn]{revtex4-1}
\pdfoutput=1
\usepackage{geometry}
\usepackage{verbatim}
\usepackage{amsmath}
\usepackage{amssymb}
\usepackage{graphicx,picture,calc}
\usepackage{hyperref}
\usepackage{braket}
\usepackage{bm}
\usepackage{epstopdf}
\usepackage[normalem]{ulem}
\usepackage[usenames,dvipsnames]{color}
\usepackage[multidot]{grffile}
\usepackage{overpic}
\usepackage{dsfont}
\newcommand{\be}{\begin{equation}}
\newcommand{\ee}{\end{equation}}
\newcommand{\bk}{{{\bf{k}}}}
\newcommand{\bp}{{{\bf{p}}}}

\newcommand{\br}{{{\bf{r}}}}

\newcommand{\bG}{{{\bf{G}}}}

\newcommand{\bq}{{\bf{q}}}

\newcommand{\bea}{\begin{eqnarray}}
\newcommand{\eea}{\end{eqnarray}}
\newcommand{\beal}{\begin{align}}
\newcommand{\eeal}{\end{align}}
\renewcommand{\i}{{\i}}

\renewcommand{\d}{{\mathrm{d}}}
\renewcommand{\i}{{\mathrm{i}}}
\newcommand{\e}{{\mathrm{e}}}

\graphicspath{{figures/compressed/}}

\begin{document}

\title{Domes of $T_c$ in single-band and multiband superconductors with finite-range attractive interactions}

\author{Nazim Boudjada}
\affiliation{Department of Physics, University of Toronto, Toronto, Ontario M5S1A7, Canada.}
\author{Finn Lasse Buessen}
\affiliation{Department of Physics, University of Toronto, Toronto, Ontario M5S1A7, Canada.}
\author{Arun Paramekanti}
\email{arunp@physics.utoronto.ca}
\affiliation{Department of Physics, University of Toronto, Toronto, Ontario M5S1A7, Canada.}

\date{\today}

\begin{abstract}
The rise and fall of the superconducting transition temperature $T_c$
upon tuning carrier density or external parameters, such as pressure or magnetic field, is ubiquitously observed
in a wide range of quantum materials.
In order to investigate such domes of $T_c$, we go beyond the prototypical attractive Hubbard model, 
and consider a lattice model of electrons coupled via instantaneous, spatially extended, attractive interactions.
By numerically solving the mean-field equations, as well as going beyond mean field theory using
a functional renormalization group approach, we find that for a characteristic interaction range $\ell$, there exists a dome in  
$T_c$ around $k_F \ell \! \sim \! {\cal O}(1)$. For multiband systems, our mean field theory shows the presence of
additional domes in the vicinity of Lifshitz transitions. Our results hold in both two and three 
dimensions and can be intuitively understood from the geometric relation between the Fermi 
surface and the interaction range. Our model may be relevant for domes of $T_c$ in 
dilute weakly coupled superconductors or in engineered cold atom systems. 
\end{abstract}

\maketitle

%%%%%%%%%%%%%%%%%%%%%%%%%%%%%%%%%%%%%%%
% Introduction
%%%%%%%%%%%%%%%%%%%%%%%%%%%%%%%%%%%%%%%

%{\it Introduction.--}
%
%A characteristic dependence of superconducting phases on the carrier density or on external parameters, such as pressure or magnetic field, is universally observed across a wide range of quantum materials, resulting in distinctive dome-like shapes in their phase diagrams. 
\section{Introduction}

Domes in the superconducting (SC) transition temperature $T_c$, observed across a broad range of quantum materials,
 typically reflect some form of underlying dynamical competition in the electronic fluid.
In heavy fermion compounds~\cite{Stewart_RMP1984,QSi_Science2010,WirthSteglich_Nature2016} and
iron pnictide materials~\cite{Wen_ARCMP2011,Si_Nature2016,Fernandes_2016}, for instance, the SC dome emerges around magnetic or nematic quantum critical points (QCPs).
In SrTiO$_3$, SC domes may possibly be driven by proximity to a ferroelectric QCP~\cite{GASTIASORO2020168107,PhysRevMaterials.3.091401,gastiasoro2020anisotropic,PhysRevB.100.094504,PhysRevB.98.024521,PhysRevB.100.094504,PhysRevLett.115.247002,Ahadieaaw0120,PhysRevB.98.104505,Neaton2019,atkinson2017influence,kedem_prb2016,kedem_prb2018}.
In the cuprates, even aside from the physics of QCPs, a decrease in the hole concentration can enhance spin-fluctuation mediated pair formation while simultaneously suppressing the superfluid density as a consequence of Mott physics or competing orders. This interplay yields the highest $T_c$ at an optimal doping~\cite{Anderson_RVB2004,Keimer_Nature2015}.
SC near QCPs has also been found in numerical simulations and field theory studies~\cite{Berg_Science2012,Chowdhury_PRB2015,Raghu_PRB2015,Wang_PRL2016,Torroba_PRB2017,Berg_ARCMP2019,Chowdhury2019}.
Finally, for ultracold atomic fermions, the highest $T_c$ appears near unitary scattering which marks the BCS-BEC crossover from weak to strong coupling SC~\cite{Randeria_ARCMP2014, STRINATI_BCSBEC2018}.

In this paper, we discuss a geometric picture of superconducting domes in systems with (nonretarded) finite-range attractive interactions. 
Our proposal is motivated by the following observation. 
In a system where electrons attract each other over a fixed characteristic range $\ell$ in real space, the typical momentum transfer in 
electron-electron scattering processes is $\Delta k \!\sim\! 1/\ell$. 
Thus, in dilute systems with Fermi momentum $k_F \!\ll\! \Delta k$, such interactions can efficiently scatter electrons across any two points on the Fermi surface (FS). 
In the opposite limit, however, when $k_F \!\gg \! \Delta k$, the aforementioned interactions lead to small-angle scattering, making it more 
challenging for electrons to explore the full FS.
Therefore, the phase space which is accessible in a single electron-scattering event initially increases 
with the size of the FS, before dropping at high densities when the locality of the interactions in momentum space suppresses global SC. 
Consequently, a dome-like dependence of $T_c$ on the electron density emerges around some intermediate Fermi momentum $k_F^\star$ which 
satisfies $k_F^\star \ell \!\sim \! {\cal O}(1)$.
The dome thus marks the crossover from predominantly global interactions to local interactions in momentum space.
From a real-space perspective, the highest $T_c$ occurs when the interaction range $\ell$ becomes comparable to the interparticle spacing.
Our work does not address the microscopic origin of such a pairing interaction or the length scale $\ell$, which are important issues  in their own right~\cite{PhysRevB.99.094524,PhysRevB.94.224515,PhysRevB.98.104505,Gorkov201604145,Gorkov}, but it
is reminiscent of the geometric Mott-Ioffe-Regel criterion which marks the crossover from coherent to incoherent electronic transport~\cite{Hussey_PhilMag2004}
without reference to an underlying mechanism.

This geometric picture for $T_c$ domes was initially proposed in the context of single-band superconductors \cite{Balatsky_Dome}, where 
semianalytical expressions of $T_c$ for different scattering potentials were obtained within a weak-coupling mean-field approach.
Our study expands on this previous work by presenting numerical solutions to the mean field equations for $T_c$ and the gap
over a wide range of densities and coupling strengths for single-band and multiband superconductors in two and three dimensions. 
In the multiband case, we find the emergence of multiple domes of $T_c$ as new bands get occupied with increasing density. In
addition, we use a functional renormalization group (FRG) approach to study the single-band model. 
We find that including corrections beyond mean field theory suppresses $T_c$ in the low-density regime, yet the maximum in $T_c$ persists. 
At higher densities near half-filling, 
where mean field theory predicts a finite $T_c$, the FRG approach shows that the effective low energy interactions flow 
towards strong forward-scattering, indicating a breakdown of superconductivity. We tentatively identify this breakdown of SC with the
onset of phase separation.

We envision that our results can be applied as a toy model for systems with critical modes or soft bosons which may induce long-range attractive interactions -- e.g. fermions experiencing fluctuating zero-momentum orders. We thus make some qualitative comparisons with results on dilute electron gases in 
bulk SrTiO$_3$. Previous work has also discussed how density-dependent
screening might lead to domes of $T_c$ in SrTiO$_3$ \cite{PhysRevB.94.224515,Gorkov201604145,Gorkov}. 
Models similar in spirit to our study have also previously been explored in the context of cuprates~\cite{KunYang_PRB2000}, SrTiO$_3$ 
\cite{Balatsky_Dome,kedem_prb2018},
FeSe on SrTiO$_3$~\cite{Rademaker_2016,dhlee_ARCMP2018}, and ultracold atomic fermions \cite{Parish_PRB2005}. 
Our work may also be relevant to ultracold Bose-Fermi mixtures, where $\ell$ could be set by the correlation length 
associated with the superfluid to Mott insulator transition~\cite{KunYang_PRB2008}.
We emphasize, however, that the $T_c$ dome we uncover is not inherently a strong-coupling phenomenon so that mean
field theory and our (truncated) FRG approach are expected to provide genuine insight.

%%%%%%%%%%%%%%%%%%%%%%%%%%%%%%%%%%%%%%%
% Model
%%%%%%%%%%%%%%%%%%%%%%%%%%%%%%%%%%%%%%%

\section{Model Hamiltonian}
\begin{figure}[t] %figure shifted for better placement
	\centering
	\begin{overpic}[width=\linewidth]{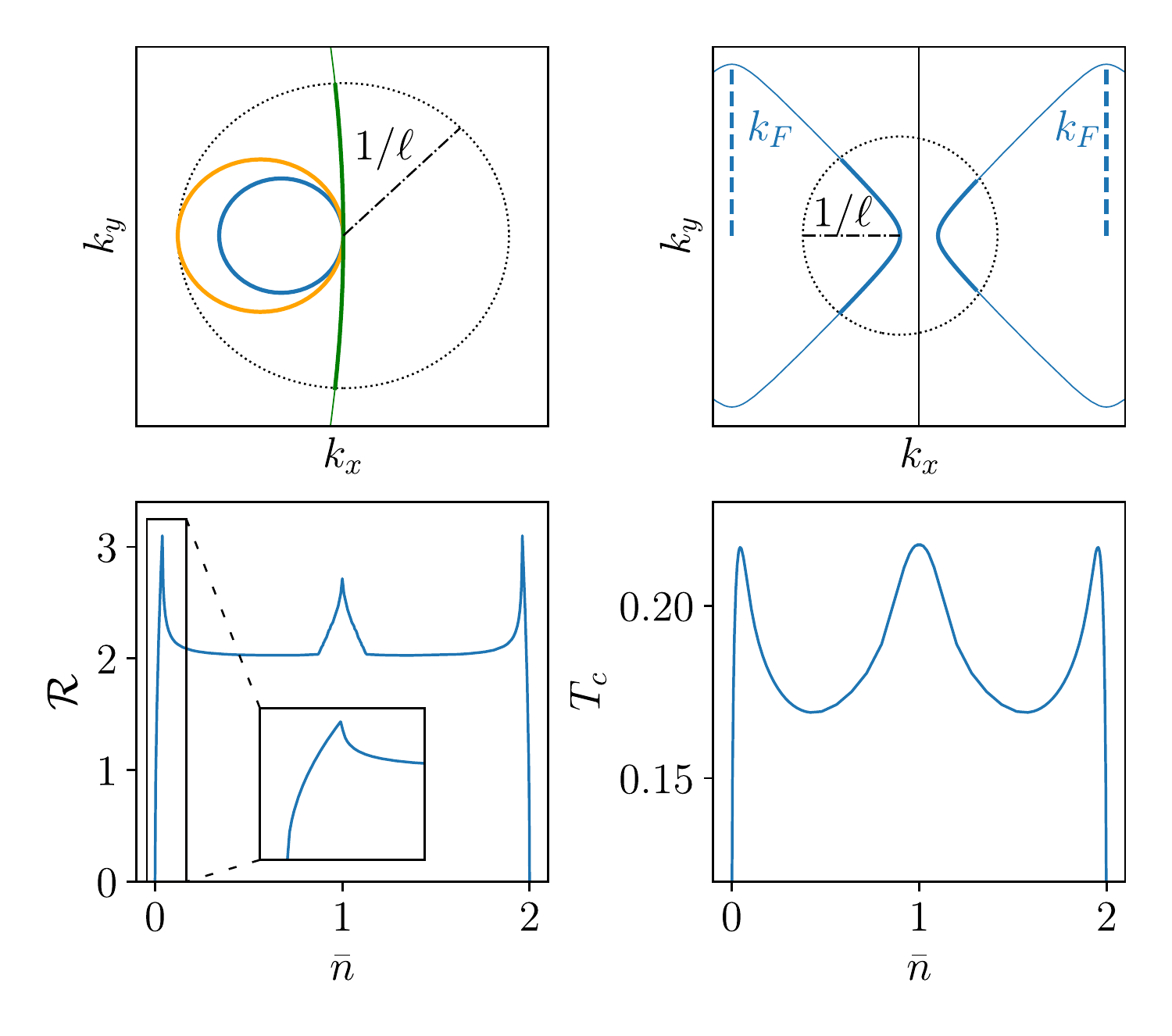}\put(6,47){(c)}\put(6,86){(a)}\put(54,47){(d)}\put(54,86){(b)}\end{overpic}
	\caption{(a) Schematic picture showing a reference momentum point on different FSs of increasing sizes (blue, orange, green). Thick lines indicate the geometrically accessible parts of the FS within the scattering circle of radius $1/\ell$.
	(b) Same as previous panel, but for larger Fermi surfaces (blue) where Umklapp scattering becomes relevant near the BZ boundary. 
	(c) Arc length ${\cal R}(\bar{n})$ of the accessible part of the FS as a function of electron density $\bar{n}$, plotted for $\ell=1$. Inset: $\mathcal{R}$ at low density showing a scaling $\sim \sqrt{\bar{n}}$.
	(d) $T_c$ as a function of $\bar{n}$ for a 2D square lattice for a fixed interaction range 
	$\ell=5$, showing peaks at the densities $\bar{n}\!\approx\!0.04,1,1.96$ (see text for details).
	}
	\label{fig:geometry}
\end{figure}
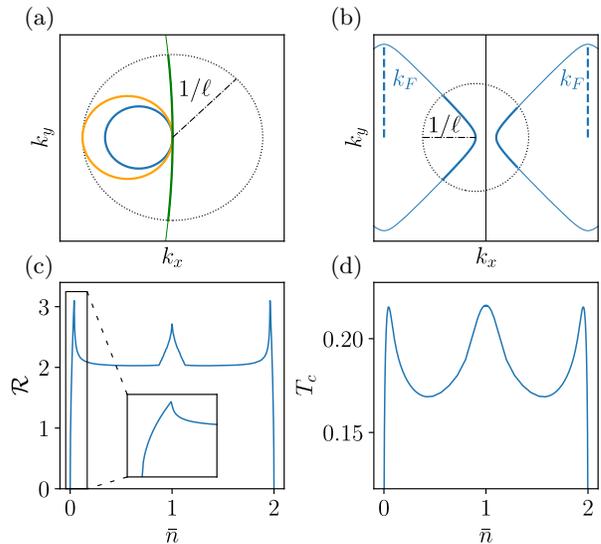

We consider a tight-binding Hamiltonian parametrized as
\begin{equation}
\mathcal{H}_0=\sum_{\bk\mu\nu} c^\dagger_\mu(\bk)H^{\mu\nu}_0(\bk)c_\nu(\bk) \,,
\end{equation}
where $\mu,\nu$ stand for generic orbital and spin 
indices which give a matrix structure to the Hamiltonian $H_0$. 
The electrons are assumed to interact via an instantaneous attractive interaction
\begin{equation}
\label{eqn:interaction}
\mathcal{H}_{\mathrm{int}}=\frac{1}{2} \int\!\! \d^d\br \!\! \int \!\! \d^d\br' ~{\cal V}(\br-\br') \hat{n}(\br) \hat{n}(\br') \,,
\end{equation}
with $\hat{n}(\br) = \sum_\mu c^\dagger_\mu(\br)c_\mu(\br)$ being the density operator at position $\br$, and ${\cal V}({\br}-{\br}')  \! < \!0$ 
being the interaction potential.

Anticipating a singlet superconducting instability, we Fourier transform the interaction to momentum space and focus on the zero center of mass pairing channel, 
which leads to the effective Hamiltonian
\begin{equation}
\label{eqn:MFHamiltonian}
\!\!\!\mathcal{H}^{\rm BCS}_{\mathrm{int}}\!=\! \frac{1}{2N} \!\sum_{\bk\bk'}c^\dagger_\mu(\bk)c^\dagger_\nu(-\bk) V(\bk\!-\!\bk') c_\nu(-\bk')c_\mu(\bk') \,,
\end{equation}
where $N$ is the total number of lattice sites and summation over repeated indices is implied. 
The interaction $V(\bk-\bk')$ is the Fourier transform of ${{\cal V}(\br-\br')}$. 
We decouple the interaction, using a Hubbard-Stratonovich transformation, via complex bosonic fields $\Delta_{\mu\nu}(\bk)$ and integrate out the fermions 
(see Appendix \ref{sec:appendix:derivation} for details). 
The resulting self-consistent matrix gap equation is given by
\begin{equation}
\small
\!\!\!\Delta(\bk)\!=\! - \frac{1}{N}\! \sum_{\bk'} V(\bk\!-\!\bk') U(\bk')\frac{\tanh\!\left[\!\frac{E(\bk')}{2T}\!\right]}{2 E(\bk')} U^\dagger(\bk')\Delta(\bk') \,,
\label{eq:gapeqn}
\end{equation}
where $T$ is the temperature, $E(\bk)$ is a diagonal matrix comprising the square roots of the eigenvalues of 
${H_0(\bk)H^\dagger_0(\bk)+\Delta(\bk)\Delta^\dagger(\bk)}$, 
and $U(\bk)$ is the corresponding eigenvector matrix. 
For a one-band model this expression reduces to the familiar single-gap equation. 
We assume a Gaussian interaction 
${{\cal V}(\br) \!=\! -g_0 \e^{-|\br|^2/2\ell^2}}$, so that ${V(\bq)\!=\!- g_0(2\pi\ell^2)^{d/2}\e^{-|\bq|^2\ell^2/2}}$ 
in $d$ spatial dimensions \footnote{Properly speaking, we set $V(\bq)$ to be a periodic Gaussian, given by $V(\bq)=  -g_0(2\pi\ell^2)^{d/2} \sum_\bG \e^{-|\bq+\bG|^2\ell^2/2}$, where $\bG$ are reciprocal lattice vectors.}. 
Here $g_0\!>\! 0$ is the pairing strength, and $\ell$ sets the range of the potential in real space (in units of the lattice constant).
For $\ell \!\to\! 0$, the interaction reduces to a Hubbard model, 
while a large value of $\ell$ favors small momentum scattering.
We note that it has already been pointed out for the single-band case that the results remain qualitatively unchanged even for alternative potentials ~\cite{Balatsky_Dome}, such as a Lorentzian or a hard sphere, with a similar characteristic range $\ell$. 
To explore the full density dependence and multiband examples, we numerically solve the gap equation for $T_c$ and the momentum dependence of the 
gap $\Delta_{\mu\nu}(\bk)$. For a fixed density $\bar{n}$, we also simultaneously solve for the chemical potential.

%%%%%%%%%%%%%%%%%%%%%%%%%%%%%%%%%%%%%%%
% Single band superconductor
%%%%%%%%%%%%%%%%%%%%%%%%%%%%%%%%%%%%%%%

\section{Single band superconductor}
\begin{figure}[t]
	\centering
	\begin{overpic}[width=\linewidth]{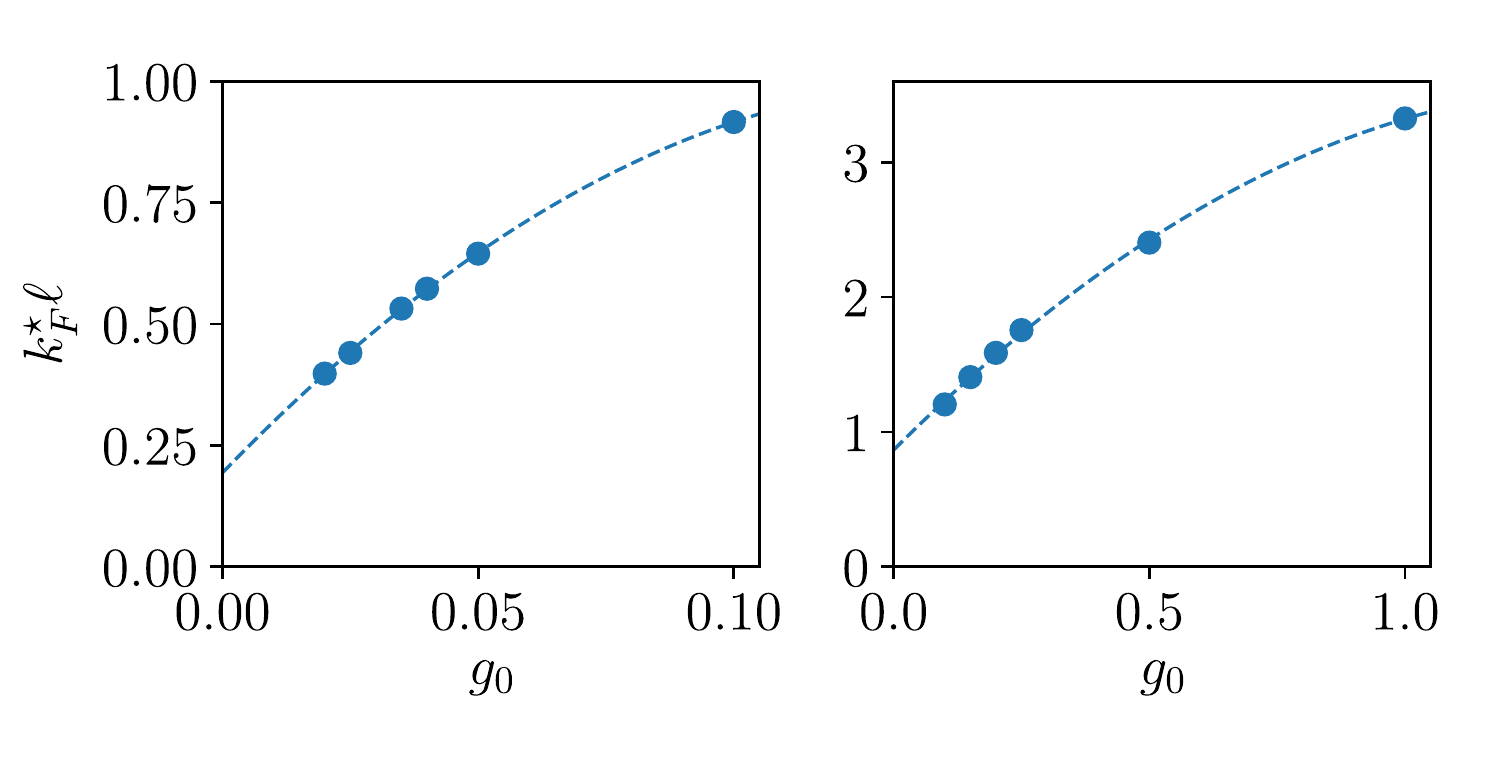}\put(15,48){(a)}\put(59,48){(b)}\end{overpic}
	\caption{Value of $k_F\ell$ at the geometric peak of $T_c$ as a function of coupling $g_0$ in (a) 2D and (b) and 3D. Extrapolation (dashed lines) to the weak-coupling limit $g_0\rightarrow0$ suggests that $k_F^\star\ell\rightarrow0.20$ in 2D and $k_F^\star\ell\rightarrow0.86$ in 3D. 
	In 3D, $k_F^\star\ell$ is larger than in 2D, indicating that the dome of $T_c$ is shifted towards lower densities, whereas the peak value of $T_c$ remains comparable in both dimensions. This is in part due to the difference in density of states at low filling.}
	\label{fig:coupling}
\end{figure}

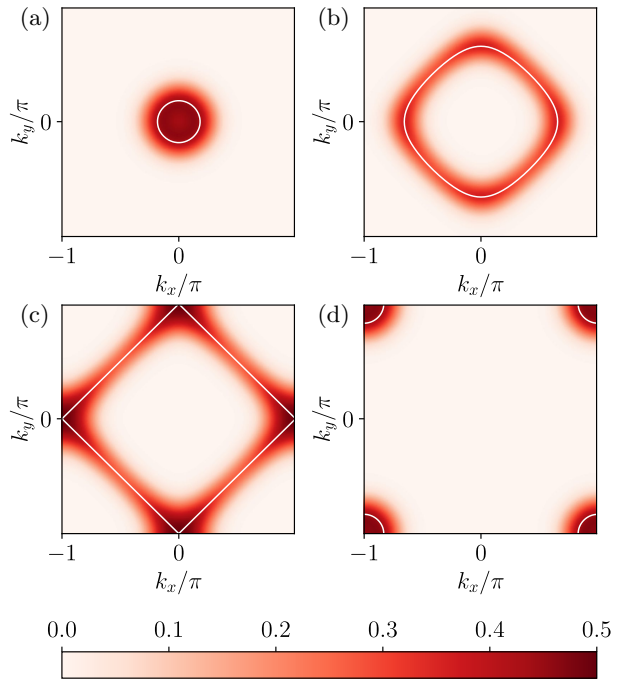
\begin{figure}[t]
	\centering
	\begin{overpic}[width=\linewidth]{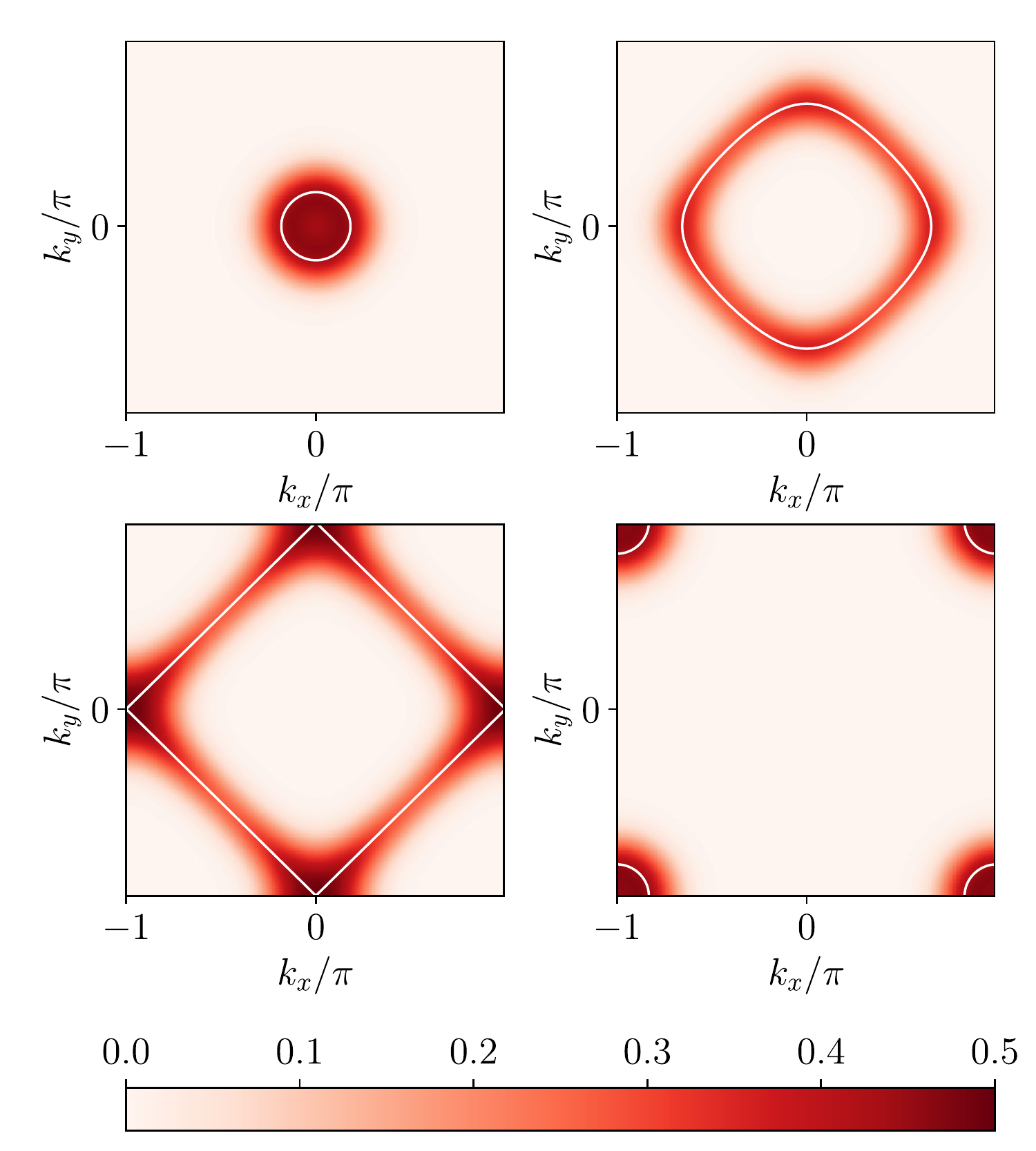}\put(5,94){(a)}\put(47,94){(b)}\put(5,52){(c)}\put(47,52){(d)}\end{overpic}
	\caption{Zero temperature solution $|\Delta(\bk)|$ (in units of $t_1=1$) to the nonlinear gap equation at different densities. (a) In the electron dilute limit ($\bar{n}=0.04$), the gap is peaked at the $\Gamma$ point while (b) at intermediate densities ($\bar{n}=0.6$) it peaks around the FS. (c) At half-filling, the gap reaches its maximum at the van Hove points ($\bar{n}=1$) and (d) in the dilute hole regime ($\bar{n}=1.96$) the maximum is at the M points.}
	\label{fig:gaps}
\end{figure}

We begin by discussing the geometric origin of domes of $T_c$ for interactions with finite range $\ell$ in a single-band
superconductor. As illustrated in Fig.~\ref{fig:geometry}(a),
for a given momentum point on the 2D FS, the arc length $\mathcal{R}$ of the FS~\footnote{We define the arc length $\cal R$ as ${\oint_{\mathrm{FS}}|\mathcal{B}_{\ell^{-1}}(\bk)\cap\mathrm{FS}|\;\d^{d-1}\bk}$ where the norm $|\cdot|$ is the geometric weight (the arc length in $d=2$ or the surface area in $d=3$) while $\mathcal{B}_{\ell^{-1}}(\bk)$ is a circle ($d=2$) or sphere $(d=3)$ of fixed radius $\ell^{-1}$ centered around a point $\bk_F$ on the Fermi surface (FS).} 
which lies within the (momentum-space) interaction range $\sim\!\! 1/\ell$ depends on the electron density.
Starting from the dilute limit, up to $k_F \ell \!=\! 1/2$, the full FS circumference $2\pi k_F$ is accessible in a single electron-scattering event [which implies a scaling $\mathcal{R}\sim\ k_F$ in the dilute limit, c.f. inset of Fig.~\ref{fig:geometry}(c)].
Beyond this value, the accessible part of the FS shrinks to $2/\ell \! < \! 2\pi k_F$ at large $k_F$. Upon further increasing the density towards 
half-filling (i.e., when $\bar{n} \!\sim\! 1$), Umklapp processes become allowed and the scattering phase space is enhanced again for 
$|\bk_F-(\bk'_F\pm\bG)|\ell=1$ where $\bG$ is a reciprocal lattice vector. This is shown pictorially in Fig.~\ref{fig:geometry}(b) 
where the accessible scattering region for a point on the left BZ includes also states from the right BZ. Finally, when ${\bar{n} \gg 1}$, 
there exists yet another geometric peak arising from small hole pockets near $(\pi,\pi)$.
Consequently, as seen in Fig.~\ref{fig:geometry}(c), the functional form of $\mathcal{R}(\bar{n})$ for a 2D square lattice with dispersion 
${\xi(\bk) \!=\!-2t_1(\cos(k_x) \!+\! \cos (k_y)) \!-\! \mu}$ and $\ell=1$ exhibits three sharp peaks: at a low density $n^\star$ corresponding 
to $k_F \ell \!=\! 1/2$, at half-filling $\bar{n}=1$, and at $2-n^\star$. Since $\mathcal{R}$ is a geometric measure of the available phase space \footnote{The available phase space for scattering is primarily set by a momentum scale cutoff $\sim 1/\ell$, instead of by an explicit energy cutoff such as the Debye frequency in conventional BCS theory.} for Cooper
pairs, we expect the peak in ${\mathcal R}$ to be reflected as a dome in $T_c$. 

Although this simple geometric argument which neglects 
the energy dependence of the density of states does not predict a peak of $T_c$ in 3D, we shall demonstrate that a smoothed version of this geometric maximum generally persists. 
To illustrate this in a simple one-band example, we compute the mean-field singlet pairing $T_c$ using $\ell\!=\! 5$ and $g_0\!=\! 1$ (in units of $t_1$), observing a peak in $T_c$ at an electron density $n^\star \!\approx\!0.04$ 
in the dilute limit as 
shown in Fig.~\ref{fig:geometry}(d).
At higher densities, $T_c$ exhibits the same additional peaks as predicted by $\mathcal{R}(\bar{n})$.
We point out that the middle peak stems from a combination of an
increased geometric overlap with states in the second BZ and an enhanced density of states near the van Hove singularity.

Our numerical solution of the gap equation shows that domes of $T_c$ also appear in 3D.
To investigate the role of the coupling strength $g_0$ in the occurrence of the dome, we compute $k_F^\star \ell$ as a function of $g_0$, where $k_F^\star$ is the angle-averaged Fermi wave vector associated with the density $n^\star$~\footnote{We define $k_F^\star = (2\pi n^\star)^{1/2}$ in 2D, and $k_F^\star = (3\pi^2 n^\star)^{1/3}$ in 3D}.
The results for the 2D square lattice as well as for the 3D cubic lattice show that the dome shifts towards smaller densities as
$g_0$ is reduced, see Fig.~\ref{fig:coupling}. However, we emphasize that the dome persists at a nonzero density even in the very weak coupling limit. 
Indeed, extrapolating our results to $g_0 \!\to \! 0$, we find a finite value $k_F^\star \ell \! \approx \! 0.20$ in 2D and $k_F^\star \ell \! \approx \! 0.86$ in 3D.
At the same time, the ratio of the critical temperature and the Fermi energy $T^\star_c/\epsilon^\star_F$ at the geometric peak remains 
moderate (with $T^\star_c/\epsilon^\star_F \!<\! 1$ for $g_0=1$, and decreasing for smaller $g_0$), implying that the dome is not a strong-coupling phenomenon.

When $k_F \ell \!\gg\! 1$, we note that many angular momentum pairing channels become quasidegenerate as seen from the 
eigenfunctions of the subleading instabilities in the linearized mean-field gap equation (see Appendix \ref{sec:appendix:eigenfunctions}). However, the dominant instability is for an ``s-wave'' gap with no nodes. Similar results were found in the context of superconductivity mediated by charge-density wave \cite{chubokov_prb2015} and nematic quantum critical points \cite{kivelson_prl2015,chubokov_npj2019}.

In Fig. \ref{fig:gaps}, we plot the gaps obtained by solving the nonlinear gap equation at $T=0$ for various densities. 
In the low electron density [panel (a)] or hole density [panel (d)] regimes, the gap peaks at the center of the small electron FS or small hole FS (i.e., at the $\Gamma$ and M points 
respectively). At intermediate densities, however, the weight of the gap is distributed on the FS in the radial direction, decaying $\sim 1/\ell$ away from $\bk_F$. In addition,
the gap shows a subdominant modulation over the FS, being larger along the $k_x=0$ and $k_y=0$ lines than along the zone diagonals.
This modulation becomes more apparent near half-filling as the FS approaches the van Hove singularity.

%%%%%%%%%%%%%%%%%%%%%%%%%%%%%%%%%%%%%%%
% Multiband
%%%%%%%%%%%%%%%%%%%%%%%%%%%%%%%%%%%%%%%

\section{Multiband superconductor}

We next generalize our mean-field results to multiband examples and demonstrate that the geometric interpretation of domes still holds. 
Furthermore, we shall see that in multiband systems, it is possible to obtain multiple domes of $T_c$ as new FSs appear with increasing density. 
To this end, we consider the two-orbital model
\begin{equation}
\!\!\!\mathcal{H}_0\!\!=\!\!\sum_{\bk\sigma} \! \begin{pmatrix}
a^\dagger_{\bk\sigma} & \! b^\dagger_{\bk\sigma}
\end{pmatrix}\!\!
\begin{pmatrix}
\xi_{a}(\bk)+V_0 & \! \delta \\
\delta & \! \xi_{b}(\bk)-V_0
\end{pmatrix}\!\!
\begin{pmatrix}
a_{\bk\sigma} \\
b_{\bk\sigma}
\end{pmatrix},
\label{eqn:twoorbitals}
\end{equation}
where
${\xi_a(\bk) \!=\! -2t_1\cos (k_x)\!-\! 2t_2\cos (k_y)\!-\! \mu}$ and
${\xi_b(\bk)\!=\!-2t_2\cos (k_x)\!-\!2t_1\cos (k_y)\!-\!\mu}$, with $\delta$ and $V_0$ being the momentum-independent interorbital 
hybridization and potential difference respectively.

\begin{figure}
	\centering
	\begin{overpic}[width=\linewidth]{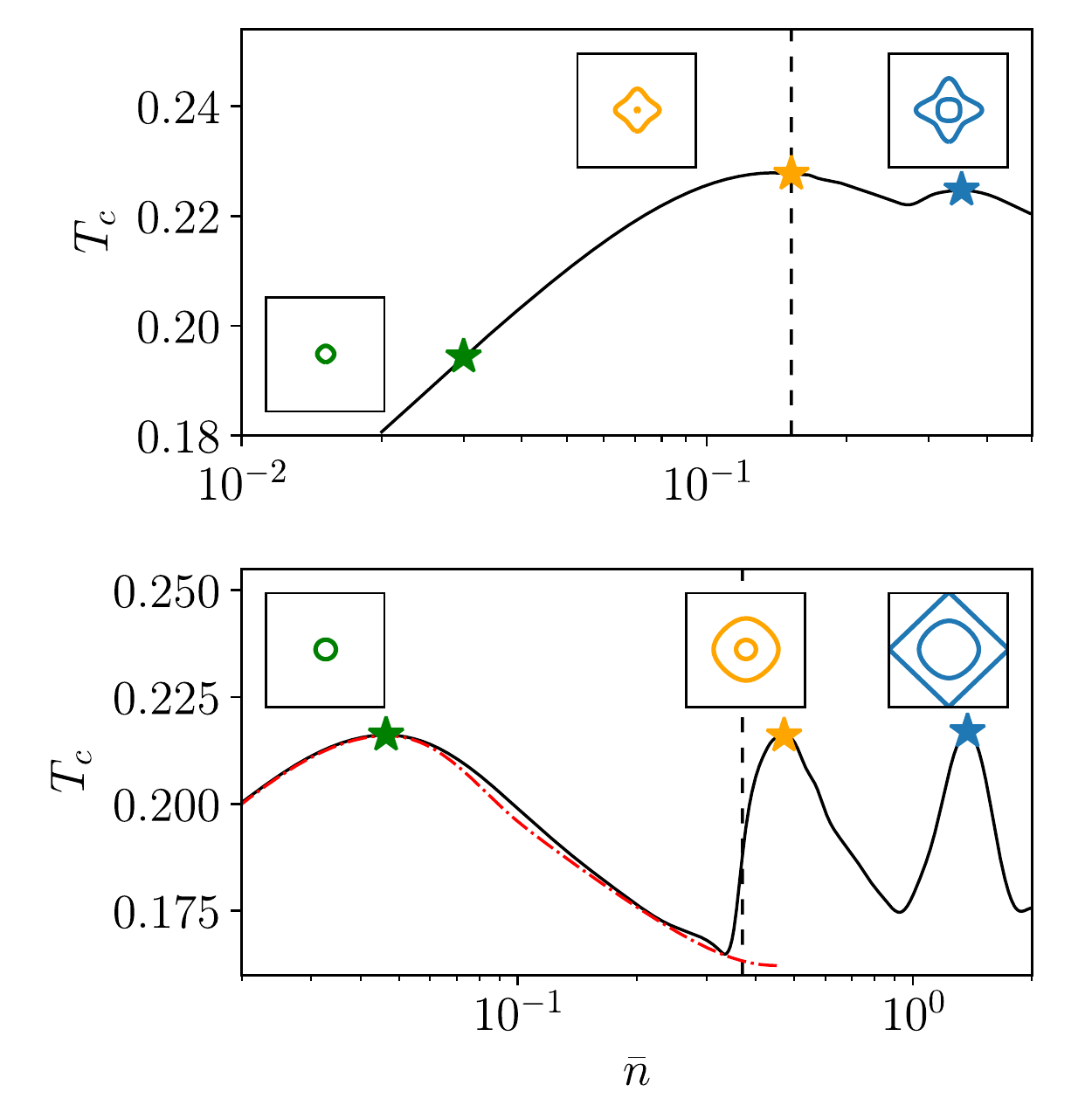}\put(3,50){(b)}\put(3,95){(a)}\end{overpic}
	\caption{Superconducting transition temperature
		$T_c$ for a two-orbital model showing a double geometric peak in the low-density regime. Colored stars correspond to the FSs in the insets. 
		Vertical dashed line marks the Lifshitz transition when the second band appears at the Fermi level.
		(a) Two hybridized elliptical FSs (b) two circular FSs with a finite potential difference (black curve). For comparison, we also show the result
		from a single orbital model (red curve) which reproduces the first peak in $T_c$.}
	\label{fig:Tc2Bands}
\end{figure}

To study the impact of interorbital hybridization, we set $t_1\!=\!1$, $\ell\!=\!5$, and $g_0\!=\!1$, and choose, for illustrative purposes, $t_2 \! = \! \delta \!=\! 0.2$. This
choice corresponds to two elliptical bands which hybridize to produce $C_4$ symmetric FSs.
In the low-density regime ($\bar{n}\ll 0.1$), only one band crosses the Fermi level and the physics is analogous to the single orbital model, i.e., a geometric dome forms at a density corresponding to the optimal value of $k_F\ell$ [orange star in Fig.~\ref{fig:Tc2Bands}(a)]. 
The parameters are chosen such that the Lifshitz transition, i.e., the appearance of the second band at the Fermi level, occurs near the maximum of the dome.
The second band then gives rise to a second geometric peak at a slightly higher density (blue star), yielding an overall double peak structure.
We note, however, that for different parameter choices, the Lifshitz transition does not necessarily coincide 
with the first peak in $T_c$: For example, increasing $\ell$ pushes the geometric peak to lower densities 
[so that $k_F^\star\ell\sim\mathcal{O}(1)$], but has no impact on the Lifshitz transition point.

Next, we study the impact of a finite potential difference between the two orbitals, keeping $\delta=0$ and setting $t_2=V_0=1$. This corresponds to two $C_4$ symmetric
bands separated in energy by $2V_0$. The two orbitals are only coupled via the constraint of the total density being fixed.
In Fig. \ref{fig:Tc2Bands}(b), we plot $T_c$ with the Fermi surfaces at the three peak densities in electron-doped regime shown in the insets, and the dashed red curve showing the one-orbital model with the same parameters. In the low-density regime, the two curves agree exactly, showing that only the lower band dictates $T_c$: A first geometric peak (marked by a green star) is reached at $\bar{n}=n^\star\approx0.04$ which corresponds to $k_F^\star\ell\approx2.85$. Near the Lifshitz transition of the two-band model 
(vertical dashed curve), the one-orbital and two-orbital models start to diverge since the higher energy band crosses the Fermi level and starts to contribute to $T_c$. 
A second geometric peak (orange star) is reached precisely when the Fermi wavevector of the new band
is such that $k_F^\star\ell\approx2.85$. At higher densities, the lower energy band reaches the van Hove point and we see the corresponding van Hove peak (blue star). As 
in the case of the one-orbital model, 
particle-hole symmetry dictates $T_c(\bar{n})=T_c(4-\bar{n})$ and an exact copy of the three peaks is obtained in the hole-doped regime $\bar{n}>2$.

%%%%%%%%%%%%%%%%%%%%%%%%%%%%%%%%%%%%%%%
% Functional RG
%%%%%%%%%%%%%%%%%%%%%%%%%%%%%%%%%%%%%%%

\section{Functional RG approach}
\begin{figure}
	\centering
	\includegraphics[width=\linewidth]{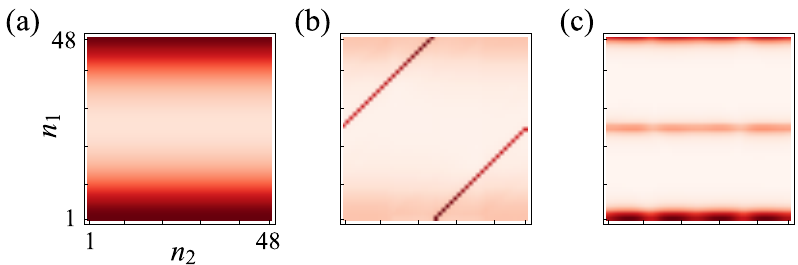}
	\caption{Effective interaction in the patching approximation. Normalized color code shows the value of the flowing interaction vertex $u_T(n_1,n_2,1)$, where 
		$(n_1,n_2)$ enumerate momentum patches around the FS. (a) Density $\bar{n}\!=\!0.17$ at $T\!=\!T_\mathrm{max}$, (b) $\bar{n}\!=\!0.17$ at $T\!=\!T_\mathrm{min}$, 
		(c) $\bar{n}\!=\!0.94$ at $T\!=\!T_\mathrm{min}$.}
	\label{fig:vertex}
\end{figure}

In deriving the gap equation Eq.~\eqref{eq:gapeqn}, we have explicitly assumed Cooper pair formation in the singlet channel. 
While yielding a structurally simple, self-consistent mean-field theory, the 
decoupling comes at the price of being inherently biased to favor the specific type of superconductivity encoded in the ansatz, potentially 
neglecting any competing phases. For $k_F \ell \!\gg\! 1$, as pointed out earlier, different patches on the FS could effectively decouple as
many angular momentum pairing channels become quasidegenerate. Furthermore, attractive interactions
could make the system unstable towards phase separation. Such effects can lead to a breakdown of
coherent superconductivity.

To investigate this breakdown -- or, conversely, justify the mean-field ansatz -- we employ an FRG approach which treats all competing interaction channels on equal footing~\cite{Metzner2012,Gersch2008,Eberlein2014}.
The resulting FRG flow equations, which relate the bare interaction as defined in Eq.~\eqref{eqn:interaction} to an effective low-energy theory by continuously tracing its evolution under infinitesimal reductions of the temperature~\cite{Honerkamp2001}, naturally have a more complex structure than the self-consistent mean-field equation, and in general cannot be solved exactly. 
For weak coupling, however, it is sufficient to include only the one-loop contributions to the flow equations for the two-particle interaction, neglecting higher-order processes~\cite{Halboth2000}, and to treat the interaction vertex in a momentum patching approximation which resolves the angular component of the momentum dependence around the FS.
In this way, a finite set of differential equations is obtained which can be solved numerically to determine the effective interaction vertex $u_T(n_1,n_2,n_3)$, 
where the $n_i$ enumerate the momentum patches around the FS.
The specific choice of momentum patches, as well as the detailed FRG flow equations, are outlined in Appendix \ref{sec:appendix:frg}. We have benchmarked our 
FRG calculations using the attractive Hubbard model.

\begin{figure}
	\centering
	\includegraphics[width=\linewidth]{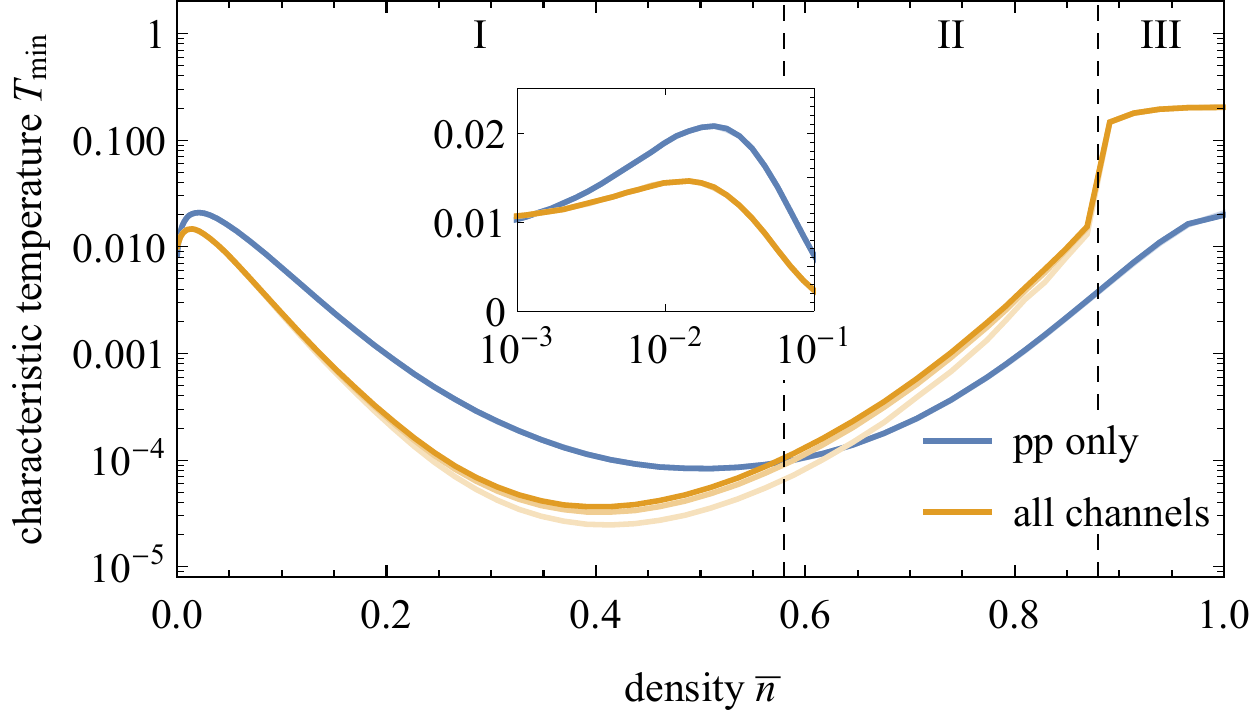}
	\caption{Characteristic temperature $T_\mathrm{min}$ for the Gaussian interaction potential with $\ell = 1$, as determined from FRG calculations which include 
	only particle-particle scattering (blue curve) or all interaction channels (orange curve). Curves plotted in opaque colors are computed using $N_p=96$ momentum 
	patches, curves in lighter colors are for $N_p=72$ and $N_p=48$. In regimes I and II, the effective vertex at $T_\mathrm{min}$ 
	captures superconducting pairing, while in regime III the interaction becomes increasingly localized in momentum space leading to a breakdown of SC.
	Inset shows the geometric dome of $T_c$ at low density.}
	\label{fig:phasediagramFRG}
\end{figure}

We now discuss our FRG calculations on the single-band model with finite range interactions for fixed ${\ell\!=\!1}$ and $g_0\!=\!\frac{3}{2\pi}$, 
while varying the density $\bar{n}$ to assess the role of competing interaction channels. 
The RG flow is initialized at an upper temperature $T_\mathrm{max}=4t_1$, which is comparable to the bandwidth, and stopped at a temperature scale 
$T_\mathrm{min}$ when the maximum component of the vertex exceeds $18t_1$, which is large compared to the bandwidth. The onset of strong
interactions at $T_\mathrm{min}$ can then be related to a putative phase transition
\footnote{The parameters are chosen such that they reproduce the superconducting $T_c$ for the single-band Hubbard model, see Appendix E}. 
We point out that the attractive Gaussian model we consider here with $\ell=1$ is similar to an extended Hubbard model, and a peak of $T_c$ is indeed observed in both models~\cite{Micnas_1988}.

In the dilute limit, $k_F \ell \ll 1$, the bare interaction at $T=T_\mathrm{max}$ has negligible momentum dependence on the FS.
The Gaussian profile becomes visible only at slightly larger $\bar{n}$ as seen in Fig.~\ref{fig:vertex}(a). 
The effective low-temperature vertex, however, for a wide range of $\bar{n}$ is dominated by a pronounced interaction between 
momentum patches which lie on opposite sides of the FS. The sign structure of this effective low-energy action is uniformly attractive, which indicates impending zero-momentum Cooper pair formation and s-wave superconductivity~\footnote{The classification of phases in terms of the low-energy vertex structure is discussed e.g. in Ref.~\cite{Metzner2012}}, see Fig.~\ref{fig:vertex}(b). 
However, at large densities ${\bar{n}>0.88}$, the initial Gaussian profile of the bare interaction 
sharpens throughout the RG flow as shown in Fig.~\ref{fig:vertex}(c), so that the renormalized $\ell \!\to\! \infty$, and forward scattering gets
enhanced. 
The isolated limit of forward scattering has previously been shown to cause an instability of the Fermi liquid~\cite{kedem2019}. 
Here, the breakdown of Cooper pairing and coherent superconductivity occurs as the low energy limit of a realistic finite range model over an extended regime of electron densities. 

We study the role of additional interaction channels in the breakdown by comparing the full FRG calculations with reduced flow 
equations that only include the particle-particle forward scattering as also captured by the mean-field ansatz.
We find that we can divide the FRG phase diagram shown in Fig.~\ref{fig:phasediagramFRG} into three regimes. 
In regime~I ($\bar{n}<0.58$) the superconducting $T_c$ is suppressed by fluctuations in additional interaction channels. Nevertheless, as shown
in the inset to Fig.~\ref{fig:phasediagramFRG}, the full FRG calculation yields a dome of $T_c$, in qualitative agreement with the simplified mean-field approach.
In regime~II ($0.58<\bar{n}<0.88$) on the other hand, unlike what is seen for the attractive Hubbard model, the finite-range character of the interactions leads to
an enhancement of $T_\mathrm{min}$ by the additional interaction channels.
Finally, in regime~III ($\bar{n}>0.88$), mean-field theory formally yields a finite $T_c$, while the full FRG approach reveals the breakdown of 
superconductivity. We tentatively identify this regime, where the renormalized $\ell \! \to \! \infty$, with phase separation
induced by extended attractive interactions.

%%%%%%%%%%%%%%%%%%%%%%%%%%%%%%%%%%%%%%%
% Discussion
%%%%%%%%%%%%%%%%%%%%%%%%%%%%%%%%%%%%%%%

\section{Conclusion}
We have provided a geometric phase space argument for the formation of $T_c$ domes in systems with spatially extended interactions.
We have shown that for multiband systems a scenario with two or more domes can arise naturally. 
In order to apply this picture to 3D bulk SrTiO$_3$, we note that the first dome with maximum transition temperature $T_c \!\approx\!0.2$\,K is 
centered at a density $\bar{n} \!\approx\! 1.2 \times 10^{18}\;{\rm cm}^{-3}$ with a Fermi energy $\epsilon_F\!\approx\! 2$\,meV. 
Demanding $k_F \ell \!\sim\! 1$ at the center of the dome yields $\ell \! \sim\! 30$\,\AA,
while requiring $T_c/\epsilon_F \! \sim \! 10^{-2}$ at this point fixes $g_0 \!\sim\! 4.5$\,meV. 
The inferred length scale $\ell$ may reflect the range of attractive interactions between polaron quasiparticles which have been reported in 
bulk SrTiO$_3$ \cite{Swartz_PNAS2018} and its interfaces
\cite{Baumberger_NMat2016, Strocov_NComms2016}.
The microscopic theory of SC of such dilute polarons remains an open issue.
%The rapid drop in $T_c$ at electron densities beyond the dome due to dominant forward scattering, as captured by the FRG, suggests that it may be possible in 
%this regime to enhance $T_c$ by impurity scattering. 
%Such scattering may allow electrons to explore different parts of the Fermi surface, and suppress the competing nonzero angular momentum pairing channels. 
It would be interesting to explore such $T_c$ domes in a wider range of experimental systems including atomic Bose-Fermi mixtures, and to
extend the FRG results by incorporating the frequency dependence of the interaction vertex. Such studies may also shed light on the interplay of 
spatially extended interactions with retardation effects in driving SC near QCPs.

%%%%%%%%%%%%%%%%%%%%%%%%%%%%%%%%%%%%%%%
% Acknowledgments
%%%%%%%%%%%%%%%%%%%%%%%%%%%%%%%%%%%%%%%

\begin{acknowledgments}
We thank M. M. Scherer and Y. Kedem for discussions. This work was funded by NSERC of Canada and FRQNT of Quebec.
The numerical simulations were performed on the JURECA cluster at the Forschungszentrum Juelich, 
and on the Cedar and Niagara clusters enabled by support provided by Compute
Ontario, SciNet, Westgrid, and Compute Canada.
SciNet is funded by:
the Canada Foundation for Innovation; the Government
of Ontario; Ontario Research Fund - Research Excellence;
and the University of Toronto.
\end{acknowledgments}

%%%%%%%%%%%%%%%%%%%%%%%%%%%%%%%%%%%%%%%
% Bibliography
%%%%%%%%%%%%%%%%%%%%%%%%%%%%%%%%%%%%%%%

\bibliography{dome}

\begin{thebibliography}{59}%
\makeatletter
\providecommand \@ifxundefined [1]{%
 \@ifx{#1\undefined}
}%
\providecommand \@ifnum [1]{%
 \ifnum #1\expandafter \@firstoftwo
 \else \expandafter \@secondoftwo
 \fi
}%
\providecommand \@ifx [1]{%
 \ifx #1\expandafter \@firstoftwo
 \else \expandafter \@secondoftwo
 \fi
}%
\providecommand \natexlab [1]{#1}%
\providecommand \enquote  [1]{``#1''}%
\providecommand \bibnamefont  [1]{#1}%
\providecommand \bibfnamefont [1]{#1}%
\providecommand \citenamefont [1]{#1}%
\providecommand \href@noop [0]{\@secondoftwo}%
\providecommand \href [0]{\begingroup \@sanitize@url \@href}%
\providecommand \@href[1]{\@@startlink{#1}\@@href}%
\providecommand \@@href[1]{\endgroup#1\@@endlink}%
\providecommand \@sanitize@url [0]{\catcode `\\12\catcode `\$12\catcode
  `\&12\catcode `\#12\catcode `\^12\catcode `\_12\catcode `\%12\relax}%
\providecommand \@@startlink[1]{}%
\providecommand \@@endlink[0]{}%
\providecommand \url  [0]{\begingroup\@sanitize@url \@url }%
\providecommand \@url [1]{\endgroup\@href {#1}{\urlprefix }}%
\providecommand \urlprefix  [0]{URL }%
\providecommand \Eprint [0]{\href }%
\providecommand \doibase [0]{http://dx.doi.org/}%
\providecommand \selectlanguage [0]{\@gobble}%
\providecommand \bibinfo  [0]{\@secondoftwo}%
\providecommand \bibfield  [0]{\@secondoftwo}%
\providecommand \translation [1]{[#1]}%
\providecommand \BibitemOpen [0]{}%
\providecommand \bibitemStop [0]{}%
\providecommand \bibitemNoStop [0]{.\EOS\space}%
\providecommand \EOS [0]{\spacefactor3000\relax}%
\providecommand \BibitemShut  [1]{\csname bibitem#1\endcsname}%
\let\auto@bib@innerbib\@empty
%</preamble>
\bibitem [{\citenamefont {Stewart}(1984)}]{Stewart_RMP1984}%
  \BibitemOpen
  \bibfield  {author} {\bibinfo {author} {\bibfnamefont {G.~R.}\ \bibnamefont
  {Stewart}},\ }\bibfield  {title} {{\color{Gray}\small \bibinfo {title}
  {Heavy-fermion systems},\ }}\href {\doibase 10.1103/RevModPhys.56.755}
  {\bibfield  {journal} {\bibinfo  {journal} {Rev. Mod. Phys.}\ }\textbf
  {\bibinfo {volume} {56}},\ \bibinfo {pages} {755} (\bibinfo {year}
  {1984})}\BibitemShut {NoStop}%
\bibitem [{\citenamefont {Si}\ and\ \citenamefont
  {Steglich}(2010)}]{QSi_Science2010}%
  \BibitemOpen
  \bibfield  {author} {\bibinfo {author} {\bibfnamefont {Q.}~\bibnamefont
  {Si}}\ and\ \bibinfo {author} {\bibfnamefont {F.}~\bibnamefont {Steglich}},\
  }\bibfield  {title} {{\color{Gray}\small \bibinfo {title} {Heavy fermions and
  quantum phase transitions},\ }}\href {\doibase 10.1126/science.1191195}
  {\bibfield  {journal} {\bibinfo  {journal} {Science}\ }\textbf {\bibinfo
  {volume} {329}},\ \bibinfo {pages} {1161} (\bibinfo {year}
  {2010})}\BibitemShut {NoStop}%
\bibitem [{\citenamefont {Wirth}\ and\ \citenamefont
  {Steglich}(2016)}]{WirthSteglich_Nature2016}%
  \BibitemOpen
  \bibfield  {author} {\bibinfo {author} {\bibfnamefont {S.}~\bibnamefont
  {Wirth}}\ and\ \bibinfo {author} {\bibfnamefont {F.}~\bibnamefont
  {Steglich}},\ }\bibfield  {title} {{\color{Gray}\small \bibinfo {title}
  {Exploring heavy fermions from macroscopic to microscopic length scales},\
  }}\href {\doibase 10.1038/natrevmats.2016.51} {\bibfield  {journal} {\bibinfo
   {journal} {Nature Reviews Materials}\ }\textbf {\bibinfo {volume} {1}},\
  \bibinfo {pages} {16051} (\bibinfo {year} {2016})}\BibitemShut {NoStop}%
\bibitem [{\citenamefont {Wen}\ and\ \citenamefont {Li}(2011)}]{Wen_ARCMP2011}%
  \BibitemOpen
  \bibfield  {author} {\bibinfo {author} {\bibfnamefont {H.-H.}\ \bibnamefont
  {Wen}}\ and\ \bibinfo {author} {\bibfnamefont {S.}~\bibnamefont {Li}},\
  }\bibfield  {title} {{\color{Gray}\small \bibinfo {title} {Materials and
  novel superconductivity in iron pnictide superconductors},\ }}\href {\doibase
  10.1146/annurev-conmatphys-062910-140518} {\bibfield  {journal} {\bibinfo
  {journal} {Annual Review of Condensed Matter Physics}\ }\textbf {\bibinfo
  {volume} {2}},\ \bibinfo {pages} {121} (\bibinfo {year} {2011})}\BibitemShut
  {NoStop}%
\bibitem [{\citenamefont {Si}\ \emph {et~al.}(2016)\citenamefont {Si},
  \citenamefont {Yu},\ and\ \citenamefont {Abrahams}}]{Si_Nature2016}%
  \BibitemOpen
  \bibfield  {author} {\bibinfo {author} {\bibfnamefont {Q.}~\bibnamefont
  {Si}}, \bibinfo {author} {\bibfnamefont {R.}~\bibnamefont {Yu}}, \ and\
  \bibinfo {author} {\bibfnamefont {E.}~\bibnamefont {Abrahams}},\ }\bibfield
  {title} {{\color{Gray}\small \bibinfo {title} {High-temperature
  superconductivity in iron pnictides and chalcogenides},\ }}\href {\doibase
  10.1038/natrevmats.2016.17} {\bibfield  {journal} {\bibinfo  {journal}
  {Nature Reviews Materials}\ }\textbf {\bibinfo {volume} {1}},\ \bibinfo
  {pages} {16017} (\bibinfo {year} {2016})}\BibitemShut {NoStop}%
\bibitem [{\citenamefont {Fernandes}\ and\ \citenamefont
  {Chubukov}(2016)}]{Fernandes_2016}%
  \BibitemOpen
  \bibfield  {author} {\bibinfo {author} {\bibfnamefont {R.~M.}\ \bibnamefont
  {Fernandes}}\ and\ \bibinfo {author} {\bibfnamefont {A.~V.}\ \bibnamefont
  {Chubukov}},\ }\bibfield  {title} {{\color{Gray}\small \bibinfo {title}
  {Low-energy microscopic models for iron-based superconductors: a review},\
  }}\href {\doibase 10.1088/1361-6633/80/1/014503} {\bibfield  {journal}
  {\bibinfo  {journal} {Reports on Progress in Physics}\ }\textbf {\bibinfo
  {volume} {80}},\ \bibinfo {pages} {014503} (\bibinfo {year}
  {2016})}\BibitemShut {NoStop}%
\bibitem [{\citenamefont {Gastiasoro}\ \emph {et~al.}(2020)\citenamefont
  {Gastiasoro}, \citenamefont {Ruhman},\ and\ \citenamefont
  {Fernandes}}]{GASTIASORO2020168107}%
  \BibitemOpen
  \bibfield  {author} {\bibinfo {author} {\bibfnamefont {M.~N.}\ \bibnamefont
  {Gastiasoro}}, \bibinfo {author} {\bibfnamefont {J.}~\bibnamefont {Ruhman}},
  \ and\ \bibinfo {author} {\bibfnamefont {R.~M.}\ \bibnamefont {Fernandes}},\
  }\bibfield  {title} {{\color{Gray}\small \bibinfo {title} {Superconductivity
  in dilute {SrTiO$_3$}: A review},\ }}\href {\doibase
  https://doi.org/10.1016/j.aop.2020.168107} {\bibfield  {journal} {\bibinfo
  {journal} {Annals of Physics}\ ,\ \bibinfo {pages} {168107}} (\bibinfo {year}
  {2020})}\BibitemShut {NoStop}%
\bibitem [{\citenamefont {Russell}\ \emph {et~al.}(2019)\citenamefont
  {Russell}, \citenamefont {Ratcliff}, \citenamefont {Ahadi}, \citenamefont
  {Dong}, \citenamefont {Stemmer},\ and\ \citenamefont
  {Harter}}]{PhysRevMaterials.3.091401}%
  \BibitemOpen
  \bibfield  {author} {\bibinfo {author} {\bibfnamefont {R.}~\bibnamefont
  {Russell}}, \bibinfo {author} {\bibfnamefont {N.}~\bibnamefont {Ratcliff}},
  \bibinfo {author} {\bibfnamefont {K.}~\bibnamefont {Ahadi}}, \bibinfo
  {author} {\bibfnamefont {L.}~\bibnamefont {Dong}}, \bibinfo {author}
  {\bibfnamefont {S.}~\bibnamefont {Stemmer}}, \ and\ \bibinfo {author}
  {\bibfnamefont {J.~W.}\ \bibnamefont {Harter}},\ }\bibfield  {title}
  {{\color{Gray}\small \bibinfo {title} {Ferroelectric enhancement of
  superconductivity in compressively strained {SrTiO}$_{3}$ films},\ }}\href
  {\doibase 10.1103/PhysRevMaterials.3.091401} {\bibfield  {journal} {\bibinfo
  {journal} {Phys. Rev. Materials}\ }\textbf {\bibinfo {volume} {3}},\ \bibinfo
  {pages} {091401} (\bibinfo {year} {2019})}\BibitemShut {NoStop}%
\bibitem [{\citenamefont {Gastiasoro}\ \emph {et~al.}()\citenamefont
  {Gastiasoro}, \citenamefont {Trevisan},\ and\ \citenamefont
  {Fernandes}}]{gastiasoro2020anisotropic}%
  \BibitemOpen
  \bibfield  {author} {\bibinfo {author} {\bibfnamefont {M.~N.}\ \bibnamefont
  {Gastiasoro}}, \bibinfo {author} {\bibfnamefont {T.~V.}\ \bibnamefont
  {Trevisan}}, \ and\ \bibinfo {author} {\bibfnamefont {R.~M.}\ \bibnamefont
  {Fernandes}},\ }\bibfield  {title} {{\color{Gray}\small \bibinfo {title}
  {Anisotropic superconductivity mediated by ferroelectric fluctuations in
  cubic systems with spin-orbit coupling},\ }}\href@noop {} {\ }\Eprint
  {http://arxiv.org/abs/2001.04919} {arXiv:2001.04919} \BibitemShut {NoStop}%
\bibitem [{\citenamefont {Kanasugi}\ and\ \citenamefont
  {Yanase}(2019)}]{PhysRevB.100.094504}%
  \BibitemOpen
  \bibfield  {author} {\bibinfo {author} {\bibfnamefont {S.}~\bibnamefont
  {Kanasugi}}\ and\ \bibinfo {author} {\bibfnamefont {Y.}~\bibnamefont
  {Yanase}},\ }\bibfield  {title} {{\color{Gray}\small \bibinfo {title}
  {Multiorbital ferroelectric superconductivity in doped {SrTiO}$_{3}$},\
  }}\href {\doibase 10.1103/PhysRevB.100.094504} {\bibfield  {journal}
  {\bibinfo  {journal} {Phys. Rev. B}\ }\textbf {\bibinfo {volume} {100}},\
  \bibinfo {pages} {094504} (\bibinfo {year} {2019})}\BibitemShut {NoStop}%
\bibitem [{\citenamefont {Kanasugi}\ and\ \citenamefont
  {Yanase}(2018)}]{PhysRevB.98.024521}%
  \BibitemOpen
  \bibfield  {author} {\bibinfo {author} {\bibfnamefont {S.}~\bibnamefont
  {Kanasugi}}\ and\ \bibinfo {author} {\bibfnamefont {Y.}~\bibnamefont
  {Yanase}},\ }\bibfield  {title} {{\color{Gray}\small \bibinfo {title}
  {Spin-orbit-coupled ferroelectric superconductivity},\ }}\href {\doibase
  10.1103/PhysRevB.98.024521} {\bibfield  {journal} {\bibinfo  {journal} {Phys.
  Rev. B}\ }\textbf {\bibinfo {volume} {98}},\ \bibinfo {pages} {024521}
  (\bibinfo {year} {2018})}\BibitemShut {NoStop}%
\bibitem [{\citenamefont {Edge}\ \emph {et~al.}(2015)\citenamefont {Edge},
  \citenamefont {Kedem}, \citenamefont {Aschauer}, \citenamefont {Spaldin},\
  and\ \citenamefont {Balatsky}}]{PhysRevLett.115.247002}%
  \BibitemOpen
  \bibfield  {author} {\bibinfo {author} {\bibfnamefont {J.~M.}\ \bibnamefont
  {Edge}}, \bibinfo {author} {\bibfnamefont {Y.}~\bibnamefont {Kedem}},
  \bibinfo {author} {\bibfnamefont {U.}~\bibnamefont {Aschauer}}, \bibinfo
  {author} {\bibfnamefont {N.~A.}\ \bibnamefont {Spaldin}}, \ and\ \bibinfo
  {author} {\bibfnamefont {A.~V.}\ \bibnamefont {Balatsky}},\ }\bibfield
  {title} {{\color{Gray}\small \bibinfo {title} {Quantum critical origin of the
  superconducting dome in {SrTiO}$_{3}$},\ }}\href {\doibase
  10.1103/PhysRevLett.115.247002} {\bibfield  {journal} {\bibinfo  {journal}
  {Phys. Rev. Lett.}\ }\textbf {\bibinfo {volume} {115}},\ \bibinfo {pages}
  {247002} (\bibinfo {year} {2015})}\BibitemShut {NoStop}%
\bibitem [{\citenamefont {Ahadi}\ \emph {et~al.}(2019)\citenamefont {Ahadi},
  \citenamefont {Galletti}, \citenamefont {Li}, \citenamefont {Salmani-Rezaie},
  \citenamefont {Wu},\ and\ \citenamefont {Stemmer}}]{Ahadieaaw0120}%
  \BibitemOpen
  \bibfield  {author} {\bibinfo {author} {\bibfnamefont {K.}~\bibnamefont
  {Ahadi}}, \bibinfo {author} {\bibfnamefont {L.}~\bibnamefont {Galletti}},
  \bibinfo {author} {\bibfnamefont {Y.}~\bibnamefont {Li}}, \bibinfo {author}
  {\bibfnamefont {S.}~\bibnamefont {Salmani-Rezaie}}, \bibinfo {author}
  {\bibfnamefont {W.}~\bibnamefont {Wu}}, \ and\ \bibinfo {author}
  {\bibfnamefont {S.}~\bibnamefont {Stemmer}},\ }\bibfield  {title}
  {{\color{Gray}\small \bibinfo {title} {Enhancing superconductivity in
  {SrTiO}$_{3}$ films with strain},\ }}\href {\doibase 10.1126/sciadv.aaw0120}
  {\bibfield  {journal} {\bibinfo  {journal} {Science Advances}\ }\textbf
  {\bibinfo {volume} {5}},\ \bibinfo {pages} {eaaw0120} (\bibinfo {year}
  {2019})}\BibitemShut {NoStop}%
\bibitem [{\citenamefont {W\"olfle}\ and\ \citenamefont
  {Balatsky}(2018)}]{PhysRevB.98.104505}%
  \BibitemOpen
  \bibfield  {author} {\bibinfo {author} {\bibfnamefont {P.}~\bibnamefont
  {W\"olfle}}\ and\ \bibinfo {author} {\bibfnamefont {A.~V.}\ \bibnamefont
  {Balatsky}},\ }\bibfield  {title} {{\color{Gray}\small \bibinfo {title}
  {Superconductivity at low density near a ferroelectric quantum critical
  point: Doped {SrTiO}$_{3}$},\ }}\href {\doibase 10.1103/PhysRevB.98.104505}
  {\bibfield  {journal} {\bibinfo  {journal} {Phys. Rev. B}\ }\textbf {\bibinfo
  {volume} {98}},\ \bibinfo {pages} {104505} (\bibinfo {year}
  {2018})}\BibitemShut {NoStop}%
\bibitem [{\citenamefont {Reyes-Lillo}\ \emph {et~al.}(2019)\citenamefont
  {Reyes-Lillo}, \citenamefont {Rabe},\ and\ \citenamefont
  {Neaton}}]{Neaton2019}%
  \BibitemOpen
  \bibfield  {author} {\bibinfo {author} {\bibfnamefont {S.~E.}\ \bibnamefont
  {Reyes-Lillo}}, \bibinfo {author} {\bibfnamefont {K.~M.}\ \bibnamefont
  {Rabe}}, \ and\ \bibinfo {author} {\bibfnamefont {J.~B.}\ \bibnamefont
  {Neaton}},\ }\bibfield  {title} {{\color{Gray}\small \bibinfo {title}
  {Ferroelectricity in [111]-oriented epitaxially strained {SrTiO}$_{3}$ from
  first principles},\ }}\href {\doibase 10.1103/PhysRevMaterials.3.030601}
  {\bibfield  {journal} {\bibinfo  {journal} {Phys. Rev. Materials}\ }\textbf
  {\bibinfo {volume} {3}},\ \bibinfo {pages} {030601} (\bibinfo {year}
  {2019})}\BibitemShut {NoStop}%
\bibitem [{\citenamefont {Atkinson}\ \emph {et~al.}(2017)\citenamefont
  {Atkinson}, \citenamefont {Lafleur},\ and\ \citenamefont
  {Raslan}}]{atkinson2017influence}%
  \BibitemOpen
  \bibfield  {author} {\bibinfo {author} {\bibfnamefont {W.~A.}\ \bibnamefont
  {Atkinson}}, \bibinfo {author} {\bibfnamefont {P.}~\bibnamefont {Lafleur}}, \
  and\ \bibinfo {author} {\bibfnamefont {A.}~\bibnamefont {Raslan}},\
  }\bibfield  {title} {{\color{Gray}\small \bibinfo {title} {Influence of the
  ferroelectric quantum critical point on {SrTiO}$_{3}$ interfaces},\ }}\href
  {\doibase 10.1103/PhysRevB.95.054107} {\bibfield  {journal} {\bibinfo
  {journal} {Phys. Rev. B}\ }\textbf {\bibinfo {volume} {95}},\ \bibinfo
  {pages} {054107} (\bibinfo {year} {2017})}\BibitemShut {NoStop}%
\bibitem [{\citenamefont {Kedem}\ \emph {et~al.}(2016)\citenamefont {Kedem},
  \citenamefont {Zhu},\ and\ \citenamefont {Balatsky}}]{kedem_prb2016}%
  \BibitemOpen
  \bibfield  {author} {\bibinfo {author} {\bibfnamefont {Y.}~\bibnamefont
  {Kedem}}, \bibinfo {author} {\bibfnamefont {J.-X.}\ \bibnamefont {Zhu}}, \
  and\ \bibinfo {author} {\bibfnamefont {A.~V.}\ \bibnamefont {Balatsky}},\
  }\bibfield  {title} {{\color{Gray}\small \bibinfo {title} {Unusual
  superconducting isotope effect in the presence of a quantum criticality},\
  }}\href {\doibase 10.1103/PhysRevB.93.184507} {\bibfield  {journal} {\bibinfo
   {journal} {Phys. Rev. B}\ }\textbf {\bibinfo {volume} {93}},\ \bibinfo
  {pages} {184507} (\bibinfo {year} {2016})}\BibitemShut {NoStop}%
\bibitem [{\citenamefont {Kedem}(2018)}]{kedem_prb2018}%
  \BibitemOpen
  \bibfield  {author} {\bibinfo {author} {\bibfnamefont {Y.}~\bibnamefont
  {Kedem}},\ }\bibfield  {title} {{\color{Gray}\small \bibinfo {title} {Novel
  pairing mechanism for superconductivity at a vanishing level of doping driven
  by critical ferroelectric modes},\ }}\href {\doibase
  10.1103/PhysRevB.98.220505} {\bibfield  {journal} {\bibinfo  {journal} {Phys.
  Rev. B}\ }\textbf {\bibinfo {volume} {98}},\ \bibinfo {pages} {220505}
  (\bibinfo {year} {2018})}\BibitemShut {NoStop}%
\bibitem [{\citenamefont {Anderson}\ \emph {et~al.}(2004)\citenamefont
  {Anderson}, \citenamefont {Lee}, \citenamefont {Randeria}, \citenamefont
  {Rice}, \citenamefont {Trivedi},\ and\ \citenamefont
  {Zhang}}]{Anderson_RVB2004}%
  \BibitemOpen
  \bibfield  {author} {\bibinfo {author} {\bibfnamefont {P.~W.}\ \bibnamefont
  {Anderson}}, \bibinfo {author} {\bibfnamefont {P.~A.}\ \bibnamefont {Lee}},
  \bibinfo {author} {\bibfnamefont {M.}~\bibnamefont {Randeria}}, \bibinfo
  {author} {\bibfnamefont {T.~M.}\ \bibnamefont {Rice}}, \bibinfo {author}
  {\bibfnamefont {N.}~\bibnamefont {Trivedi}}, \ and\ \bibinfo {author}
  {\bibfnamefont {F.~C.}\ \bibnamefont {Zhang}},\ }\bibfield  {title}
  {{\color{Gray}\small \bibinfo {title} {The physics behind high-temperature
  superconducting cuprates: the ~plain vanilla~ version of {RVB}},\ }}\href
  {\doibase 10.1088/0953-8984/16/24/r02} {\bibfield  {journal} {\bibinfo
  {journal} {Journal of Physics: Condensed Matter}\ }\textbf {\bibinfo {volume}
  {16}},\ \bibinfo {pages} {R755} (\bibinfo {year} {2004})}\BibitemShut
  {NoStop}%
\bibitem [{\citenamefont {Keimer}\ \emph {et~al.}(2015)\citenamefont {Keimer},
  \citenamefont {Kivelson}, \citenamefont {Norman}, \citenamefont {Uchida},\
  and\ \citenamefont {Zaanen}}]{Keimer_Nature2015}%
  \BibitemOpen
  \bibfield  {author} {\bibinfo {author} {\bibfnamefont {B.}~\bibnamefont
  {Keimer}}, \bibinfo {author} {\bibfnamefont {S.~A.}\ \bibnamefont
  {Kivelson}}, \bibinfo {author} {\bibfnamefont {M.~R.}\ \bibnamefont
  {Norman}}, \bibinfo {author} {\bibfnamefont {S.}~\bibnamefont {Uchida}}, \
  and\ \bibinfo {author} {\bibfnamefont {J.}~\bibnamefont {Zaanen}},\
  }\bibfield  {title} {{\color{Gray}\small \bibinfo {title} {From quantum
  matter to high-temperature superconductivity in copper oxides},\ }}\href
  {\doibase 10.1038/nature14165} {\bibfield  {journal} {\bibinfo  {journal}
  {Nature}\ }\textbf {\bibinfo {volume} {518}},\ \bibinfo {pages} {179}
  (\bibinfo {year} {2015})}\BibitemShut {NoStop}%
\bibitem [{\citenamefont {Berg}\ \emph {et~al.}(2012)\citenamefont {Berg},
  \citenamefont {Metlitski},\ and\ \citenamefont {Sachdev}}]{Berg_Science2012}%
  \BibitemOpen
  \bibfield  {author} {\bibinfo {author} {\bibfnamefont {E.}~\bibnamefont
  {Berg}}, \bibinfo {author} {\bibfnamefont {M.~A.}\ \bibnamefont {Metlitski}},
  \ and\ \bibinfo {author} {\bibfnamefont {S.}~\bibnamefont {Sachdev}},\
  }\bibfield  {title} {{\color{Gray}\small \bibinfo {title} {Sign-problem-free
  {Quantum} {Monte} {Carlo} of the onset of antiferromagnetism in metals},\
  }}\href {\doibase 10.1126/science.1227769} {\bibfield  {journal} {\bibinfo
  {journal} {Science}\ }\textbf {\bibinfo {volume} {338}},\ \bibinfo {pages}
  {1606} (\bibinfo {year} {2012})}\BibitemShut {NoStop}%
\bibitem [{\citenamefont {Chowdhury}\ and\ \citenamefont
  {Sachdev}(2015)}]{Chowdhury_PRB2015}%
  \BibitemOpen
  \bibfield  {author} {\bibinfo {author} {\bibfnamefont {D.}~\bibnamefont
  {Chowdhury}}\ and\ \bibinfo {author} {\bibfnamefont {S.}~\bibnamefont
  {Sachdev}},\ }\bibfield  {title} {{\color{Gray}\small \bibinfo {title} {Higgs
  criticality in a two-dimensional metal},\ }}\href {\doibase
  10.1103/PhysRevB.91.115123} {\bibfield  {journal} {\bibinfo  {journal} {Phys.
  Rev. B}\ }\textbf {\bibinfo {volume} {91}},\ \bibinfo {pages} {115123}
  (\bibinfo {year} {2015})}\BibitemShut {NoStop}%
\bibitem [{\citenamefont {Raghu}\ \emph {et~al.}(2015)\citenamefont {Raghu},
  \citenamefont {Torroba},\ and\ \citenamefont {Wang}}]{Raghu_PRB2015}%
  \BibitemOpen
  \bibfield  {author} {\bibinfo {author} {\bibfnamefont {S.}~\bibnamefont
  {Raghu}}, \bibinfo {author} {\bibfnamefont {G.}~\bibnamefont {Torroba}}, \
  and\ \bibinfo {author} {\bibfnamefont {H.}~\bibnamefont {Wang}},\ }\bibfield
  {title} {{\color{Gray}\small \bibinfo {title} {Metallic quantum critical
  points with finite {BCS} couplings},\ }}\href {\doibase
  10.1103/PhysRevB.92.205104} {\bibfield  {journal} {\bibinfo  {journal} {Phys.
  Rev. B}\ }\textbf {\bibinfo {volume} {92}},\ \bibinfo {pages} {205104}
  (\bibinfo {year} {2015})}\BibitemShut {NoStop}%
\bibitem [{\citenamefont {Wang}\ \emph
  {et~al.}(2016{\natexlab{a}})\citenamefont {Wang}, \citenamefont {Abanov},
  \citenamefont {Altshuler}, \citenamefont {Yuzbashyan},\ and\ \citenamefont
  {Chubukov}}]{Wang_PRL2016}%
  \BibitemOpen
  \bibfield  {author} {\bibinfo {author} {\bibfnamefont {Y.}~\bibnamefont
  {Wang}}, \bibinfo {author} {\bibfnamefont {A.}~\bibnamefont {Abanov}},
  \bibinfo {author} {\bibfnamefont {B.~L.}\ \bibnamefont {Altshuler}}, \bibinfo
  {author} {\bibfnamefont {E.~A.}\ \bibnamefont {Yuzbashyan}}, \ and\ \bibinfo
  {author} {\bibfnamefont {A.~V.}\ \bibnamefont {Chubukov}},\ }\bibfield
  {title} {{\color{Gray}\small \bibinfo {title} {Superconductivity near a
  quantum-critical point: The special role of the first {Matsubara}
  frequency},\ }}\href {\doibase 10.1103/PhysRevLett.117.157001} {\bibfield
  {journal} {\bibinfo  {journal} {Phys. Rev. Lett.}\ }\textbf {\bibinfo
  {volume} {117}},\ \bibinfo {pages} {157001} (\bibinfo {year}
  {2016}{\natexlab{a}})}\BibitemShut {NoStop}%
\bibitem [{\citenamefont {Wang}\ \emph {et~al.}(2017)\citenamefont {Wang},
  \citenamefont {Raghu},\ and\ \citenamefont {Torroba}}]{Torroba_PRB2017}%
  \BibitemOpen
  \bibfield  {author} {\bibinfo {author} {\bibfnamefont {H.}~\bibnamefont
  {Wang}}, \bibinfo {author} {\bibfnamefont {S.}~\bibnamefont {Raghu}}, \ and\
  \bibinfo {author} {\bibfnamefont {G.}~\bibnamefont {Torroba}},\ }\bibfield
  {title} {{\color{Gray}\small \bibinfo {title} {Non-{Fermi}-liquid
  superconductivity: Eliashberg approach versus the renormalization group},\
  }}\href {\doibase 10.1103/PhysRevB.95.165137} {\bibfield  {journal} {\bibinfo
   {journal} {Phys. Rev. B}\ }\textbf {\bibinfo {volume} {95}},\ \bibinfo
  {pages} {165137} (\bibinfo {year} {2017})}\BibitemShut {NoStop}%
\bibitem [{\citenamefont {Berg}\ \emph {et~al.}(2019)\citenamefont {Berg},
  \citenamefont {Lederer}, \citenamefont {Schattner},\ and\ \citenamefont
  {Trebst}}]{Berg_ARCMP2019}%
  \BibitemOpen
  \bibfield  {author} {\bibinfo {author} {\bibfnamefont {E.}~\bibnamefont
  {Berg}}, \bibinfo {author} {\bibfnamefont {S.}~\bibnamefont {Lederer}},
  \bibinfo {author} {\bibfnamefont {Y.}~\bibnamefont {Schattner}}, \ and\
  \bibinfo {author} {\bibfnamefont {S.}~\bibnamefont {Trebst}},\ }\bibfield
  {title} {{\color{Gray}\small \bibinfo {title} {Monte {Carlo} studies of
  quantum critical metals},\ }}\href {\doibase
  10.1146/annurev-conmatphys-031218-013339} {\bibfield  {journal} {\bibinfo
  {journal} {Annual Review of Condensed Matter Physics}\ }\textbf {\bibinfo
  {volume} {10}},\ \bibinfo {pages} {63} (\bibinfo {year} {2019})}\BibitemShut
  {NoStop}%
\bibitem [{\citenamefont {{Chowdhury}}\ and\ \citenamefont
  {{Berg}}()}]{Chowdhury2019}%
  \BibitemOpen
  \bibfield  {author} {\bibinfo {author} {\bibfnamefont {D.}~\bibnamefont
  {{Chowdhury}}}\ and\ \bibinfo {author} {\bibfnamefont {E.}~\bibnamefont
  {{Berg}}},\ }\bibfield  {title} {{\color{Gray}\small \bibinfo {title} {{The
  unreasonable effectiveness of Eliashberg theory for pairing of non-Fermi
  liquids}},\ }}\href@noop {} {\ }\Eprint {http://arxiv.org/abs/1912.07646}
  {arXiv:1912.07646} \BibitemShut {NoStop}%
\bibitem [{\citenamefont {Randeria}\ and\ \citenamefont
  {Taylor}(2014)}]{Randeria_ARCMP2014}%
  \BibitemOpen
  \bibfield  {author} {\bibinfo {author} {\bibfnamefont {M.}~\bibnamefont
  {Randeria}}\ and\ \bibinfo {author} {\bibfnamefont {E.}~\bibnamefont
  {Taylor}},\ }\bibfield  {title} {{\color{Gray}\small \bibinfo {title}
  {Crossover from {Bardeen}-{Cooper}-{Schrieffer} to {Bose}-{Einstein}
  condensation and the unitary {Fermi} gas},\ }}\href {\doibase
  10.1146/annurev-conmatphys-031113-133829} {\bibfield  {journal} {\bibinfo
  {journal} {Annual Review of Condensed Matter Physics}\ }\textbf {\bibinfo
  {volume} {5}},\ \bibinfo {pages} {209} (\bibinfo {year} {2014})}\BibitemShut
  {NoStop}%
\bibitem [{\citenamefont {Strinati}\ \emph {et~al.}(2018)\citenamefont
  {Strinati}, \citenamefont {Pieri}, \citenamefont {Röpke}, \citenamefont
  {Schuck},\ and\ \citenamefont {Urban}}]{STRINATI_BCSBEC2018}%
  \BibitemOpen
  \bibfield  {author} {\bibinfo {author} {\bibfnamefont {G.~C.}\ \bibnamefont
  {Strinati}}, \bibinfo {author} {\bibfnamefont {P.}~\bibnamefont {Pieri}},
  \bibinfo {author} {\bibfnamefont {G.}~\bibnamefont {Röpke}}, \bibinfo
  {author} {\bibfnamefont {P.}~\bibnamefont {Schuck}}, \ and\ \bibinfo {author}
  {\bibfnamefont {M.}~\bibnamefont {Urban}},\ }\bibfield  {title}
  {{\color{Gray}\small \bibinfo {title} {The {BCS}-{BEC} crossover: From
  ultra-cold {Fermi} gases to nuclear systems},\ }}\href {\doibase
  https://doi.org/10.1016/j.physrep.2018.02.004} {\bibfield  {journal}
  {\bibinfo  {journal} {Physics Reports}\ }\textbf {\bibinfo {volume} {738}},\
  \bibinfo {pages} {1 } (\bibinfo {year} {2018})}\BibitemShut {NoStop}%
\bibitem [{\citenamefont {Gastiasoro}\ \emph {et~al.}(2019)\citenamefont
  {Gastiasoro}, \citenamefont {Chubukov},\ and\ \citenamefont
  {Fernandes}}]{PhysRevB.99.094524}%
  \BibitemOpen
  \bibfield  {author} {\bibinfo {author} {\bibfnamefont {M.~N.}\ \bibnamefont
  {Gastiasoro}}, \bibinfo {author} {\bibfnamefont {A.~V.}\ \bibnamefont
  {Chubukov}}, \ and\ \bibinfo {author} {\bibfnamefont {R.~M.}\ \bibnamefont
  {Fernandes}},\ }\bibfield  {title} {{\color{Gray}\small \bibinfo {title}
  {Phonon-mediated superconductivity in low carrier-density systems},\ }}\href
  {\doibase 10.1103/PhysRevB.99.094524} {\bibfield  {journal} {\bibinfo
  {journal} {Phys. Rev. B}\ }\textbf {\bibinfo {volume} {99}},\ \bibinfo
  {pages} {094524} (\bibinfo {year} {2019})}\BibitemShut {NoStop}%
\bibitem [{\citenamefont {Ruhman}\ and\ \citenamefont
  {Lee}(2016)}]{PhysRevB.94.224515}%
  \BibitemOpen
  \bibfield  {author} {\bibinfo {author} {\bibfnamefont {J.}~\bibnamefont
  {Ruhman}}\ and\ \bibinfo {author} {\bibfnamefont {P.~A.}\ \bibnamefont
  {Lee}},\ }\bibfield  {title} {{\color{Gray}\small \bibinfo {title}
  {Superconductivity at very low density: The case of strontium titanate},\
  }}\href {\doibase 10.1103/PhysRevB.94.224515} {\bibfield  {journal} {\bibinfo
   {journal} {Phys. Rev. B}\ }\textbf {\bibinfo {volume} {94}},\ \bibinfo
  {pages} {224515} (\bibinfo {year} {2016})}\BibitemShut {NoStop}%
\bibitem [{\citenamefont {Gor{\textquoteright}kov}(2016)}]{Gorkov201604145}%
  \BibitemOpen
  \bibfield  {author} {\bibinfo {author} {\bibfnamefont {L.~P.}\ \bibnamefont
  {Gor{\textquoteright}kov}},\ }\bibfield  {title} {{\color{Gray}\small
  \bibinfo {title} {Phonon mechanism in the most dilute superconductor n-type
  {SrTiO}$_{3}$},\ }}\href {\doibase 10.1073/pnas.1604145113} {\bibfield
  {journal} {\bibinfo  {journal} {Proceedings of the National Academy of
  Sciences}\ }\textbf {\bibinfo {volume} {113}},\ \bibinfo {pages} {4646}
  (\bibinfo {year} {2016})}\BibitemShut {NoStop}%
\bibitem [{\citenamefont {Gor{\textquoteright}kov}(2017)}]{Gorkov}%
  \BibitemOpen
  \bibfield  {author} {\bibinfo {author} {\bibfnamefont {L.~P.}\ \bibnamefont
  {Gor{\textquoteright}kov}},\ }\bibfield  {title} {{\color{Gray}\small
  \bibinfo {title} {Back to mechanisms of superconductivity in low-doped
  strontium titanate},\ }}\href {\doibase 10.1007/s10948-017-4000-1} {\bibfield
   {journal} {\bibinfo  {journal} {J Supercond Nov Magn}\ }\textbf {\bibinfo
  {volume} {30}},\ \bibinfo {pages} {845} (\bibinfo {year} {2017})}\BibitemShut
  {NoStop}%
\bibitem [{\citenamefont {Hussey}\ \emph {et~al.}(2004)\citenamefont {Hussey},
  \citenamefont {Takenaka},\ and\ \citenamefont {Takagi}}]{Hussey_PhilMag2004}%
  \BibitemOpen
  \bibfield  {author} {\bibinfo {author} {\bibfnamefont {N.~E.}\ \bibnamefont
  {Hussey}}, \bibinfo {author} {\bibfnamefont {K.}~\bibnamefont {Takenaka}}, \
  and\ \bibinfo {author} {\bibfnamefont {H.}~\bibnamefont {Takagi}},\
  }\bibfield  {title} {{\color{Gray}\small \bibinfo {title} {Universality of
  the {Mott}-{Ioffe}-{Regel} limit in metals},\ }}\href {\doibase
  10.1080/14786430410001716944} {\bibfield  {journal} {\bibinfo  {journal}
  {Philosophical Magazine}\ }\textbf {\bibinfo {volume} {84}},\ \bibinfo
  {pages} {2847} (\bibinfo {year} {2004})}\BibitemShut {NoStop}%
\bibitem [{\citenamefont {Langmann}\ \emph {et~al.}(2019)\citenamefont
  {Langmann}, \citenamefont {Triola},\ and\ \citenamefont
  {Balatsky}}]{Balatsky_Dome}%
  \BibitemOpen
  \bibfield  {author} {\bibinfo {author} {\bibfnamefont {E.}~\bibnamefont
  {Langmann}}, \bibinfo {author} {\bibfnamefont {C.}~\bibnamefont {Triola}}, \
  and\ \bibinfo {author} {\bibfnamefont {A.~V.}\ \bibnamefont {Balatsky}},\
  }\bibfield  {title} {{\color{Gray}\small \bibinfo {title} {Ubiquity of
  superconducting domes in the bardeen-cooper-schrieffer theory with
  finite-range potentials},\ }}\href {\doibase 10.1103/PhysRevLett.122.157001}
  {\bibfield  {journal} {\bibinfo  {journal} {Phys. Rev. Lett.}\ }\textbf
  {\bibinfo {volume} {122}},\ \bibinfo {pages} {157001} (\bibinfo {year}
  {2019})}\BibitemShut {NoStop}%
\bibitem [{\citenamefont {Yang}\ and\ \citenamefont
  {Sondhi}(2000)}]{KunYang_PRB2000}%
  \BibitemOpen
  \bibfield  {author} {\bibinfo {author} {\bibfnamefont {K.}~\bibnamefont
  {Yang}}\ and\ \bibinfo {author} {\bibfnamefont {S.~L.}\ \bibnamefont
  {Sondhi}},\ }\bibfield  {title} {{\color{Gray}\small \bibinfo {title}
  {Low-energy collective modes, {Ginzburg}-{Landau} theory, and pseudogap
  behavior in superconductors with long-range pairing interactions},\ }}\href
  {\doibase 10.1103/PhysRevB.62.11778} {\bibfield  {journal} {\bibinfo
  {journal} {Phys. Rev. B}\ }\textbf {\bibinfo {volume} {62}},\ \bibinfo
  {pages} {11778} (\bibinfo {year} {2000})}\BibitemShut {NoStop}%
\bibitem [{\citenamefont {Rademaker}\ \emph {et~al.}(2016)\citenamefont
  {Rademaker}, \citenamefont {Wang}, \citenamefont {Berlijn},\ and\
  \citenamefont {Johnston}}]{Rademaker_2016}%
  \BibitemOpen
  \bibfield  {author} {\bibinfo {author} {\bibfnamefont {L.}~\bibnamefont
  {Rademaker}}, \bibinfo {author} {\bibfnamefont {Y.}~\bibnamefont {Wang}},
  \bibinfo {author} {\bibfnamefont {T.}~\bibnamefont {Berlijn}}, \ and\
  \bibinfo {author} {\bibfnamefont {S.}~\bibnamefont {Johnston}},\ }\bibfield
  {title} {{\color{Gray}\small \bibinfo {title} {Enhanced superconductivity due
  to forward scattering in {FeSe} thin films on {SrTiO}$_{3}$ substrates},\
  }}\href {\doibase 10.1088/1367-2630/18/2/022001} {\bibfield  {journal}
  {\bibinfo  {journal} {New Journal of Physics}\ }\textbf {\bibinfo {volume}
  {18}},\ \bibinfo {pages} {022001} (\bibinfo {year} {2016})}\BibitemShut
  {NoStop}%
\bibitem [{\citenamefont {Lee}(2018)}]{dhlee_ARCMP2018}%
  \BibitemOpen
  \bibfield  {author} {\bibinfo {author} {\bibfnamefont {D.-H.}\ \bibnamefont
  {Lee}},\ }\bibfield  {title} {{\color{Gray}\small \bibinfo {title} {Routes to
  high-temperature superconductivity: A lesson from {FeSe}/{SrTiO}$_{3}$},\
  }}\href {\doibase 10.1146/annurev-conmatphys-033117-053942} {\bibfield
  {journal} {\bibinfo  {journal} {Annual Review of Condensed Matter Physics}\
  }\textbf {\bibinfo {volume} {9}},\ \bibinfo {pages} {261} (\bibinfo {year}
  {2018})}\BibitemShut {NoStop}%
\bibitem [{\citenamefont {Parish}\ \emph {et~al.}(2005)\citenamefont {Parish},
  \citenamefont {Mihaila}, \citenamefont {Timmermans}, \citenamefont
  {Blagoev},\ and\ \citenamefont {Littlewood}}]{Parish_PRB2005}%
  \BibitemOpen
  \bibfield  {author} {\bibinfo {author} {\bibfnamefont {M.~M.}\ \bibnamefont
  {Parish}}, \bibinfo {author} {\bibfnamefont {B.}~\bibnamefont {Mihaila}},
  \bibinfo {author} {\bibfnamefont {E.~M.}\ \bibnamefont {Timmermans}},
  \bibinfo {author} {\bibfnamefont {K.~B.}\ \bibnamefont {Blagoev}}, \ and\
  \bibinfo {author} {\bibfnamefont {P.~B.}\ \bibnamefont {Littlewood}},\
  }\bibfield  {title} {{\color{Gray}\small \bibinfo {title} {Bcs-bec crossover
  with a finite-range interaction},\ }}\href {\doibase
  10.1103/PhysRevB.71.064513} {\bibfield  {journal} {\bibinfo  {journal} {Phys.
  Rev. B}\ }\textbf {\bibinfo {volume} {71}},\ \bibinfo {pages} {064513}
  (\bibinfo {year} {2005})}\BibitemShut {NoStop}%
\bibitem [{\citenamefont {Yang}(2008)}]{KunYang_PRB2008}%
  \BibitemOpen
  \bibfield  {author} {\bibinfo {author} {\bibfnamefont {K.}~\bibnamefont
  {Yang}},\ }\bibfield  {title} {{\color{Gray}\small \bibinfo {title}
  {Superfluid-insulator transition and fermion pairing in {Bose}-{Fermi}
  mixtures},\ }}\href {\doibase 10.1103/PhysRevB.77.085115} {\bibfield
  {journal} {\bibinfo  {journal} {Phys. Rev. B}\ }\textbf {\bibinfo {volume}
  {77}},\ \bibinfo {pages} {085115} (\bibinfo {year} {2008})}\BibitemShut
  {NoStop}%
\bibitem [{Note1()}]{Note1}%
  \BibitemOpen
  \bibinfo {note} {Properly speaking, we set $V({\protect \bf {q}})$ to be a
  periodic Gaussian, given by $V({\protect \bf {q}})= -g_0(2\pi \ell ^2)^{d/2}
  \DOTSB \sum@ \slimits@ _{{\protect \bf {G}}}{\protect \mathrm
  {e}}^{-|{\protect \bf {q}}+{{\protect \bf {G}}}|^2\ell ^2/2}$, where
  ${{\protect \bf {G}}}$ are reciprocal lattice vectors.}\BibitemShut {Stop}%
\bibitem [{Note2()}]{Note2}%
  \BibitemOpen
  \bibinfo {note} {We define the arc length $\protect \cal R$ as ${\DOTSI
  \ointop \ilimits@ _{\protect \mathrm {FS}}|\protect \mathcal {B}_{\ell
  ^{-1}}({{\protect \bf {k}}})\cap \protect \mathrm {FS}|\protect \tmspace
  +\thickmuskip {.2777em}{\protect \mathrm {d}}^{d-1}{{\protect \bf {k}}}}$
  where the norm $|\cdot |$ is the geometric weight (the arc length in $d=2$ or
  the surface area in $d=3$) while $\protect \mathcal {B}_{\ell
  ^{-1}}({{\protect \bf {k}}})$ is a circle ($d=2$) or sphere $(d=3)$ of fixed
  radius $\ell ^{-1}$ centered around a point ${{\protect \bf {k}}}_F$ on the
  Fermi surface (FS).}\BibitemShut {Stop}%
\bibitem [{Note3()}]{Note3}%
  \BibitemOpen
  \bibinfo {note} {The available phase space for scattering is primarily set by
  a momentum scale cutoff $\sim 1/\ell $, instead of by an explicit energy
  cutoff such as the Debye frequency in conventional BCS theory.}\BibitemShut
  {Stop}%
\bibitem [{Note4()}]{Note4}%
  \BibitemOpen
  \bibinfo {note} {We define $k_F^\star = (2\pi n^\star )^{1/2}$ in 2D, and
  $k_F^\star = (3\pi ^2 n^\star )^{1/3}$ in 3D}\BibitemShut {NoStop}%
\bibitem [{\citenamefont {Wang}\ and\ \citenamefont
  {Chubukov}(2015)}]{chubokov_prb2015}%
  \BibitemOpen
  \bibfield  {author} {\bibinfo {author} {\bibfnamefont {Y.}~\bibnamefont
  {Wang}}\ and\ \bibinfo {author} {\bibfnamefont {A.~V.}\ \bibnamefont
  {Chubukov}},\ }\bibfield  {title} {{\color{Gray}\small \bibinfo {title}
  {Enhancement of superconductivity at the onset of charge-density-wave order
  in a metal},\ }}\href {\doibase 10.1103/PhysRevB.92.125108} {\bibfield
  {journal} {\bibinfo  {journal} {Phys. Rev. B}\ }\textbf {\bibinfo {volume}
  {92}},\ \bibinfo {pages} {125108} (\bibinfo {year} {2015})}\BibitemShut
  {NoStop}%
\bibitem [{\citenamefont {Lederer}\ \emph {et~al.}(2015)\citenamefont
  {Lederer}, \citenamefont {Schattner}, \citenamefont {Berg},\ and\
  \citenamefont {Kivelson}}]{kivelson_prl2015}%
  \BibitemOpen
  \bibfield  {author} {\bibinfo {author} {\bibfnamefont {S.}~\bibnamefont
  {Lederer}}, \bibinfo {author} {\bibfnamefont {Y.}~\bibnamefont {Schattner}},
  \bibinfo {author} {\bibfnamefont {E.}~\bibnamefont {Berg}}, \ and\ \bibinfo
  {author} {\bibfnamefont {S.~A.}\ \bibnamefont {Kivelson}},\ }\bibfield
  {title} {{\color{Gray}\small \bibinfo {title} {Enhancement of
  superconductivity near a nematic quantum critical point},\ }}\href {\doibase
  10.1103/PhysRevLett.114.097001} {\bibfield  {journal} {\bibinfo  {journal}
  {Phys. Rev. Lett.}\ }\textbf {\bibinfo {volume} {114}},\ \bibinfo {pages}
  {097001} (\bibinfo {year} {2015})}\BibitemShut {NoStop}%
\bibitem [{\citenamefont {Klein}\ \emph {et~al.}(2019)\citenamefont {Klein},
  \citenamefont {Wu},\ and\ \citenamefont {Chubukov}}]{chubokov_npj2019}%
  \BibitemOpen
  \bibfield  {author} {\bibinfo {author} {\bibfnamefont {A.}~\bibnamefont
  {Klein}}, \bibinfo {author} {\bibfnamefont {Y.}~\bibnamefont {Wu}}, \ and\
  \bibinfo {author} {\bibfnamefont {A.}~\bibnamefont {Chubukov}},\ }\bibfield
  {title} {{\color{Gray}\small \bibinfo {title} {Multiple intertwined pairing
  states and temperature-sensitive gap anisotropy for superconductivity at a
  nematic quantum-critical point},\ }}\href {\doibase
  10.1038/s41535-019-0192-x} {\bibfield  {journal} {\bibinfo  {journal} {npj
  Quantum Mater}\ }\textbf {\bibinfo {volume} {4}} (\bibinfo {year} {2019}),\
  10.1038/s41535-019-0192-x}\BibitemShut {NoStop}%
\bibitem [{\citenamefont {Metzner}\ \emph {et~al.}(2012)\citenamefont
  {Metzner}, \citenamefont {Salmhofer}, \citenamefont {Honerkamp},
  \citenamefont {Meden},\ and\ \citenamefont
  {Sch{\"{o}}nhammer}}]{Metzner2012}%
  \BibitemOpen
  \bibfield  {author} {\bibinfo {author} {\bibfnamefont {W.}~\bibnamefont
  {Metzner}}, \bibinfo {author} {\bibfnamefont {M.}~\bibnamefont {Salmhofer}},
  \bibinfo {author} {\bibfnamefont {C.}~\bibnamefont {Honerkamp}}, \bibinfo
  {author} {\bibfnamefont {V.}~\bibnamefont {Meden}}, \ and\ \bibinfo {author}
  {\bibfnamefont {K.}~\bibnamefont {Sch{\"{o}}nhammer}},\ }\bibfield  {title}
  {{\color{Gray}\small \bibinfo {title} {{Functional renormalization group
  approach to correlated fermion systems}},\ }}\href {\doibase
  10.1103/RevModPhys.84.299} {\bibfield  {journal} {\bibinfo  {journal}
  {Reviews of Modern Physics}\ }\textbf {\bibinfo {volume} {84}},\ \bibinfo
  {pages} {299} (\bibinfo {year} {2012})}\BibitemShut {NoStop}%
\bibitem [{\citenamefont {Gersch}\ \emph {et~al.}(2008)\citenamefont {Gersch},
  \citenamefont {Honerkamp},\ and\ \citenamefont {Metzner}}]{Gersch2008}%
  \BibitemOpen
  \bibfield  {author} {\bibinfo {author} {\bibfnamefont {R.}~\bibnamefont
  {Gersch}}, \bibinfo {author} {\bibfnamefont {C.}~\bibnamefont {Honerkamp}}, \
  and\ \bibinfo {author} {\bibfnamefont {W.}~\bibnamefont {Metzner}},\
  }\bibfield  {title} {{\color{Gray}\small \bibinfo {title} {{Superconductivity
  in the attractive Hubbard model: functional renormalization group
  analysis}},\ }}\href {\doibase 10.1088/1367-2630/10/4/045003} {\bibfield
  {journal} {\bibinfo  {journal} {New Journal of Physics}\ }\textbf {\bibinfo
  {volume} {10}},\ \bibinfo {pages} {045003} (\bibinfo {year}
  {2008})}\BibitemShut {NoStop}%
\bibitem [{\citenamefont {Eberlein}(2014)}]{Eberlein2014}%
  \BibitemOpen
  \bibfield  {author} {\bibinfo {author} {\bibfnamefont {A.}~\bibnamefont
  {Eberlein}},\ }\bibfield  {title} {{\color{Gray}\small \bibinfo {title}
  {{Fermionic two-loop functional renormalization group for correlated
  fermions: Method and application to the attractive Hubbard model}},\ }}\href
  {\doibase 10.1103/PhysRevB.90.115125} {\bibfield  {journal} {\bibinfo
  {journal} {Physical Review B}\ }\textbf {\bibinfo {volume} {90}},\ \bibinfo
  {pages} {115125} (\bibinfo {year} {2014})}\BibitemShut {NoStop}%
\bibitem [{\citenamefont {Honerkamp}\ and\ \citenamefont
  {Salmhofer}(2001)}]{Honerkamp2001}%
  \BibitemOpen
  \bibfield  {author} {\bibinfo {author} {\bibfnamefont {C.}~\bibnamefont
  {Honerkamp}}\ and\ \bibinfo {author} {\bibfnamefont {M.}~\bibnamefont
  {Salmhofer}},\ }\bibfield  {title} {{\color{Gray}\small \bibinfo {title}
  {{Temperature-flow renormalization group and the competition between
  superconductivity and ferromagnetism}},\ }}\href {\doibase
  10.1103/PhysRevB.64.184516} {\bibfield  {journal} {\bibinfo  {journal}
  {Physical Review B}\ }\textbf {\bibinfo {volume} {64}},\ \bibinfo {pages}
  {184516} (\bibinfo {year} {2001})}\BibitemShut {NoStop}%
\bibitem [{\citenamefont {Halboth}\ and\ \citenamefont
  {Metzner}(2000)}]{Halboth2000}%
  \BibitemOpen
  \bibfield  {author} {\bibinfo {author} {\bibfnamefont {C.~J.}\ \bibnamefont
  {Halboth}}\ and\ \bibinfo {author} {\bibfnamefont {W.}~\bibnamefont
  {Metzner}},\ }\bibfield  {title} {{\color{Gray}\small \bibinfo {title}
  {{Renormalization-group analysis of the two-dimensional Hubbard model}},\
  }}\href {\doibase 10.1103/physrevb.61.7364} {\bibfield  {journal} {\bibinfo
  {journal} {Physical Review B}\ }\textbf {\bibinfo {volume} {61}},\ \bibinfo
  {pages} {7364} (\bibinfo {year} {2000})}\BibitemShut {NoStop}%
\bibitem [{Note5()}]{Note5}%
  \BibitemOpen
  \bibinfo {note} {The parameters are chosen such that they reproduce the
  superconducting $T_c$ for the single band Hubbard model, see Appendix
  E}\BibitemShut {NoStop}%
\bibitem [{\citenamefont {Micnas}\ \emph {et~al.}(1988)\citenamefont {Micnas},
  \citenamefont {Ranninger},\ and\ \citenamefont
  {Robaszkiewicz}}]{Micnas_1988}%
  \BibitemOpen
  \bibfield  {author} {\bibinfo {author} {\bibfnamefont {R.}~\bibnamefont
  {Micnas}}, \bibinfo {author} {\bibfnamefont {J.}~\bibnamefont {Ranninger}}, \
  and\ \bibinfo {author} {\bibfnamefont {S.}~\bibnamefont {Robaszkiewicz}},\
  }\bibfield  {title} {{\color{Gray}\small \bibinfo {title} {An extended
  {Hubbard} model with inter-site attraction in two dimensions and high-{Tc}
  superconductivity},\ }}\href {\doibase 10.1088/0022-3719/21/6/009} {\bibfield
   {journal} {\bibinfo  {journal} {Journal of Physics C: Solid State Physics}\
  }\textbf {\bibinfo {volume} {21}},\ \bibinfo {pages} {L145} (\bibinfo {year}
  {1988})}\BibitemShut {NoStop}%
\bibitem [{Note6()}]{Note6}%
  \BibitemOpen
  \bibinfo {note} {The classification of phases in terms of the low-energy
  vertex structure is discussed e.g. in Ref.~\cite {Metzner2012}}\BibitemShut
  {NoStop}%
\bibitem [{\citenamefont {Kedem}(2019)}]{kedem2019}%
  \BibitemOpen
  \bibfield  {author} {\bibinfo {author} {\bibfnamefont {Y.}~\bibnamefont
  {Kedem}},\ }\href@noop {} {\bibinfo {title} {Instability of a dilute {Fermi}
  liquid in the presence of forward scattering},\ } (\bibinfo {year} {2019}),\
  \Eprint {http://arxiv.org/abs/1904.11290} {arXiv:1904.11290
  [cond-mat.supr-con]} \BibitemShut {NoStop}%
\bibitem [{\citenamefont {Swartz}\ \emph {et~al.}(2018)\citenamefont {Swartz},
  \citenamefont {Inoue}, \citenamefont {Merz}, \citenamefont {Hikita},
  \citenamefont {Raghu}, \citenamefont {Devereaux}, \citenamefont {Johnston},\
  and\ \citenamefont {Hwang}}]{Swartz_PNAS2018}%
  \BibitemOpen
  \bibfield  {author} {\bibinfo {author} {\bibfnamefont {A.~G.}\ \bibnamefont
  {Swartz}}, \bibinfo {author} {\bibfnamefont {H.}~\bibnamefont {Inoue}},
  \bibinfo {author} {\bibfnamefont {T.~A.}\ \bibnamefont {Merz}}, \bibinfo
  {author} {\bibfnamefont {Y.}~\bibnamefont {Hikita}}, \bibinfo {author}
  {\bibfnamefont {S.}~\bibnamefont {Raghu}}, \bibinfo {author} {\bibfnamefont
  {T.~P.}\ \bibnamefont {Devereaux}}, \bibinfo {author} {\bibfnamefont
  {S.}~\bibnamefont {Johnston}}, \ and\ \bibinfo {author} {\bibfnamefont
  {H.~Y.}\ \bibnamefont {Hwang}},\ }\bibfield  {title} {{\color{Gray}\small
  \bibinfo {title} {Polaronic behavior in a weak-coupling superconductor},\
  }}\href {\doibase 10.1073/pnas.1713916115} {\bibfield  {journal} {\bibinfo
  {journal} {Proceedings of the National Academy of Sciences}\ }\textbf
  {\bibinfo {volume} {115}},\ \bibinfo {pages} {1475} (\bibinfo {year}
  {2018})}\BibitemShut {NoStop}%
\bibitem [{\citenamefont {Wang}\ \emph
  {et~al.}(2016{\natexlab{b}})\citenamefont {Wang}, \citenamefont
  {McKeown~Walker}, \citenamefont {Tamai}, \citenamefont {Wang}, \citenamefont
  {Ristic}, \citenamefont {Bruno}, \citenamefont {de~la Torre}, \citenamefont
  {Ricc{\`o}}, \citenamefont {Plumb}, \citenamefont {Shi}, \citenamefont
  {Hlawenka}, \citenamefont {S{\'a}nchez-Barriga}, \citenamefont {Varykhalov},
  \citenamefont {Kim}, \citenamefont {Hoesch}, \citenamefont {King},
  \citenamefont {Meevasana}, \citenamefont {Diebold}, \citenamefont {Mesot},
  \citenamefont {Moritz}, \citenamefont {Devereaux}, \citenamefont {Radovic},\
  and\ \citenamefont {Baumberger}}]{Baumberger_NMat2016}%
  \BibitemOpen
  \bibfield  {author} {\bibinfo {author} {\bibfnamefont {Z.}~\bibnamefont
  {Wang}}, \bibinfo {author} {\bibfnamefont {S.}~\bibnamefont
  {McKeown~Walker}}, \bibinfo {author} {\bibfnamefont {A.}~\bibnamefont
  {Tamai}}, \bibinfo {author} {\bibfnamefont {Y.}~\bibnamefont {Wang}},
  \bibinfo {author} {\bibfnamefont {Z.}~\bibnamefont {Ristic}}, \bibinfo
  {author} {\bibfnamefont {F.~Y.}\ \bibnamefont {Bruno}}, \bibinfo {author}
  {\bibfnamefont {A.}~\bibnamefont {de~la Torre}}, \bibinfo {author}
  {\bibfnamefont {S.}~\bibnamefont {Ricc{\`o}}}, \bibinfo {author}
  {\bibfnamefont {N.~C.}\ \bibnamefont {Plumb}}, \bibinfo {author}
  {\bibfnamefont {M.}~\bibnamefont {Shi}}, \bibinfo {author} {\bibfnamefont
  {P.}~\bibnamefont {Hlawenka}}, \bibinfo {author} {\bibfnamefont
  {J.}~\bibnamefont {S{\'a}nchez-Barriga}}, \bibinfo {author} {\bibfnamefont
  {A.}~\bibnamefont {Varykhalov}}, \bibinfo {author} {\bibfnamefont {T.~K.}\
  \bibnamefont {Kim}}, \bibinfo {author} {\bibfnamefont {M.}~\bibnamefont
  {Hoesch}}, \bibinfo {author} {\bibfnamefont {P.~D.~C.}\ \bibnamefont {King}},
  \bibinfo {author} {\bibfnamefont {W.}~\bibnamefont {Meevasana}}, \bibinfo
  {author} {\bibfnamefont {U.}~\bibnamefont {Diebold}}, \bibinfo {author}
  {\bibfnamefont {J.}~\bibnamefont {Mesot}}, \bibinfo {author} {\bibfnamefont
  {B.}~\bibnamefont {Moritz}}, \bibinfo {author} {\bibfnamefont {T.~P.}\
  \bibnamefont {Devereaux}}, \bibinfo {author} {\bibfnamefont {M.}~\bibnamefont
  {Radovic}}, \ and\ \bibinfo {author} {\bibfnamefont {F.}~\bibnamefont
  {Baumberger}},\ }\bibfield  {title} {{\color{Gray}\small \bibinfo {title}
  {Tailoring the nature and strength of electron--phonon interactions in the
  {SrTiO}$_{3}$(001) {2D} electron liquid},\ }}\href {\doibase
  10.1038/nmat4623} {\bibfield  {journal} {\bibinfo  {journal} {Nature
  Materials}\ }\textbf {\bibinfo {volume} {15}},\ \bibinfo {pages} {835}
  (\bibinfo {year} {2016}{\natexlab{b}})}\BibitemShut {NoStop}%
\bibitem [{\citenamefont {Cancellieri}\ \emph {et~al.}(2016)\citenamefont
  {Cancellieri}, \citenamefont {Mishchenko}, \citenamefont {Aschauer},
  \citenamefont {Filippetti}, \citenamefont {Faber}, \citenamefont {Bari{\v
  s}i{\'c}}, \citenamefont {Rogalev}, \citenamefont {Schmitt}, \citenamefont
  {Nagaosa},\ and\ \citenamefont {Strocov}}]{Strocov_NComms2016}%
  \BibitemOpen
  \bibfield  {author} {\bibinfo {author} {\bibfnamefont {C.}~\bibnamefont
  {Cancellieri}}, \bibinfo {author} {\bibfnamefont {A.~S.}\ \bibnamefont
  {Mishchenko}}, \bibinfo {author} {\bibfnamefont {U.}~\bibnamefont
  {Aschauer}}, \bibinfo {author} {\bibfnamefont {A.}~\bibnamefont
  {Filippetti}}, \bibinfo {author} {\bibfnamefont {C.}~\bibnamefont {Faber}},
  \bibinfo {author} {\bibfnamefont {O.~S.}\ \bibnamefont {Bari{\v s}i{\'c}}},
  \bibinfo {author} {\bibfnamefont {V.~A.}\ \bibnamefont {Rogalev}}, \bibinfo
  {author} {\bibfnamefont {T.}~\bibnamefont {Schmitt}}, \bibinfo {author}
  {\bibfnamefont {N.}~\bibnamefont {Nagaosa}}, \ and\ \bibinfo {author}
  {\bibfnamefont {V.~N.}\ \bibnamefont {Strocov}},\ }\bibfield  {title}
  {{\color{Gray}\small \bibinfo {title} {Polaronic metal state at the
  {LaAlO}$_{3}$/{SrTiO}$_{3}$ interface},\ }}\href {\doibase
  10.1038/ncomms10386} {\bibfield  {journal} {\bibinfo  {journal} {Nature
  Communications}\ }\textbf {\bibinfo {volume} {7}},\ \bibinfo {pages} {10386}
  (\bibinfo {year} {2016})}\BibitemShut {NoStop}%
\end{thebibliography}%

%%%%%%%%%%%%%%%%%%%%%%%%%%%%%%%%%%%%%%%
%%%%%%%%%%%%%%%%%%%%%%%%%%%%%%%%%%%%%%%
% Appendix
%%%%%%%%%%%%%%%%%%%%%%%%%%%%%%%%%%%%%%%
%%%%%%%%%%%%%%%%%%%%%%%%%%%%%%%%%%%%%%%

\appendix

\begin{widetext}
	
	%%%%%%%%%%%%%%%%%%%%%%%%%%%%%%%%%%%%%%%
	% Derivation of multiband gap equation
	%%%%%%%%%%%%%%%%%%%%%%%%%%%%%%%%%%%%%%%
	
	\section{Derivation of multiband gap equation}
	\label{sec:appendix:derivation}
	We start by writing the imaginary-time action for free fermions:
	\begin{equation}
	S_0=\frac{1}{V^2}\int\d^d\br\;\int\d^d\br'\int_0^\beta\d\tau\;\bar{\psi}_\mu(\br,\tau)[\partial_\tau\delta_{\mu\nu}-H_0^{\mu\nu}(\br,\br')]\psi_\nu(\br',\tau),
	\end{equation}
	where $V=Na^d$ is the volume of the $d$-dimensional cubic system with $N$ sites of lattice constant $a$. Working in units where $a=1$, we can Fourier transform $S_0$ to momentum and Matsubara frequency space for a translationally invariant system:
	\begin{eqnarray}
	S_0&=&\frac{1}{N}\sum_{\bk\omega_n}\bar{\psi}_\mu(\bk,\i\omega_n)[\i\omega_n\delta_{\mu\nu}-H_0^{\mu\nu}(\bk)]\psi_\nu(\bk,\i\omega_n)\nonumber\\
	&=&\frac{1}{2N}\sum_{\bk\omega_n}\left(\bar{\psi}_\mu(\bk,\i\omega_n)\underbrace{[\i\omega_n\delta_{\mu\nu}-H_0^{\mu\nu}(\bk)]}_{G_{0p}^{-1}(\bk,\i\omega_n)}\psi_\nu(\bk,\i\omega_n)+\psi_\mu(-\bk,-\i\omega_n)\underbrace{[\i\omega_n\delta_{\mu\nu}+H_0^{\nu\mu}(-\bk)]}_{G_{0h}^{-1}(-\bk,-\i\omega_n)}\bar{\psi}_\nu(-\bk,-\i\omega_n)\right)\nonumber\\
	&=&\frac{1}{2N}\sum_{\bk\omega_n}
	\begin{pmatrix}\bar{\psi}(\bk,\i\omega_n) & \psi(-\bk,-\i\omega_n)\end{pmatrix}
	\begin{pmatrix}G_{0p}^{-1}(\bk,\i\omega_n) & 0 \\ 
	0 & G_{0h}^{-1}(-\bk,-\i\omega_n)\end{pmatrix}
	\begin{pmatrix}\psi(\bk,\i\omega_n) \\ \bar{\psi}(-\bk,-\i\omega_n)\end{pmatrix} \,,
	\end{eqnarray}
	where $G^{-1}_{0p}$ and $G^{-1}_{0h}$ are the matrix noninteracting Green's functions, neglecting self-energy corrections. The instantaneous interaction is, in real-space:
	\begin{equation}
	S_{\mathrm{int}}=\frac{1}{2V^2}\int\d^d\br\;\int\d^d\br'\int_0^\beta\d\tau\;\bar{\psi}_\mu(\br,\tau)\bar{\psi}_\nu(\br',\tau)\mathcal{V}(\br-\br')\psi_\nu(\br',\tau)\psi_\mu(\br,\tau) \,,
	\end{equation}
	and in $\bk$ space, if we only keep zero center of mass momentum terms:
	\begin{equation}
	S_{\mathrm{int}}=\frac{1}{2\beta N^2}\sum_{\bk\bk'}\sum_{\omega_n\omega_m}\bar{\psi}_\mu(\bk,\i\omega_n)\bar{\psi}_\nu(-\bk,-\i\omega_n)V(\bk-\bk')\psi_\nu(-\bk',-\i\omega_m)\psi_\mu(\bk',\i\omega_m) \,,
	\end{equation}
	with $V(\bk-\bk')<0$ the Fourier transform of $\mathcal{V}(\br)$. We now introduce the complex fields $\Delta_{\mu\nu}(\bk)$:
	\small
	\begin{align}
	\e^{-S_{\mathrm{int}}}\propto\int\mathcal{D}[\bar{\Delta},\Delta]\;\exp\bigg(&-\frac{1}{2N}\sum_{\bk\omega_n}\Big(\frac{\beta}{N}\sum_{\bk'}\Delta_{\mu\nu}(\bk')\Delta_{\nu\mu}^*(\bk)F(\bk-\bk') \nonumber\\
	&+\psi_\nu(-\bk,-\i\omega_n)\Delta_{\nu\mu}^*(\bk)\psi_\mu(\bk,\i\omega_n)+\bar{\psi}_\mu(\bk,\i\omega_n)\Delta_{\mu\nu}(\bk)\bar{\psi}_\nu(-\bk,-\i\omega_n)\Big)\bigg)
	\end{align}
	\normalsize
	with $F(\bk-\bk')=\frac{1}{V}\int\d^d\br\;\frac{\e^{\i(\bk-\bk')\cdot \br}}{\mathcal{V}(\br)}$. The total partition function, $\mathcal{Z}=\int\mathcal{D}[\bar{\psi},\psi]\e^{-(S_0+S_{\mathrm{int}})}$ up to normalization constants, becomes quadratic in the fermion fields:
	\begin{eqnarray}
	\hspace{-5mm}
	\mathcal{Z}&=&\int\mathcal{D}[\bar{\Delta},\Delta]\exp\left(-\frac{\beta}{2N^2}\sum_{\bk\bk'}\Delta_{\mu\nu}(\bk')\Delta_{\nu\mu}^*(\bk)F(\bk-\bk')\right)\nonumber\\
	&\times&\int\mathcal{D}[\bar{\psi},\psi]\exp\left(-\frac{1}{2N}\sum_{\bk\omega_n}
	\underbrace{\begin{pmatrix}\bar{\psi}(\bk,\i\omega_n) & \psi(-\bk,-\i\omega_n)\end{pmatrix}}_{\bar{\Psi}(\bk,\i\omega_n)}
	\underbrace{\begin{pmatrix}G_{0p}^{-1}(\bk,\i\omega_n) & \Delta(\bk) \\ 
		\Delta^\dagger(\bk) & G_{0h}^{-1}(-\bk,-\i\omega_n)\end{pmatrix}}_{\mathcal{G}^{-1}(\bk,\i\omega_n)}
	\underbrace{\begin{pmatrix}\psi(\bk,\i\omega_n) \\ \bar{\psi}(-\bk,-\i\omega_n)\end{pmatrix}}_{\Psi(\bk,\i\omega_n)}\right) \,.
	\end{eqnarray}
	We can proceed by integrating out the fermions and obtain the effective action $\mathcal{Z}=\int\mathcal{D}[\bar{\Delta},\Delta]\e^{-S_{\mathrm{eff}}}$:
	\begin{equation}
	S_{\mathrm{eff}}=\underbrace{\frac{\beta}{2N^2}\sum_{\bk\bk'}\Delta_{\mu\nu}(\bk')\Delta_{\nu\mu}^*(\bk)F(\bk-\bk')}_{S_1}+\underbrace{\frac{1}{2N}\sum_{\bk\omega_n}\mathrm{tr}\log\begin{pmatrix}\i\omega_n\mathbb{I}-H_0(\bk) & \Delta(\bk) \\ 
		\Delta^\dagger(\bk) & \i\omega_n\mathbb{I}+H_0^\mathrm{T}(-\bk)\end{pmatrix}}_{S_2},
	\end{equation}
	To obtain the equation of motion, we need to set  $\frac{\delta S_{\mathrm{eff}}}{\delta \Delta^*_{\sigma\lambda}(\bp)}=0$. Varying $S_1$ is straightforward:
	\begin{equation}
	\frac{1}{\beta}\frac{\delta S_1}{\delta \Delta^*_{\sigma\lambda}(\bp)}=\frac{1}{2N^2}\sum_{\bk\bk'}\delta_{\nu\sigma}\delta_{\mu\lambda}\delta^{(d)}(\bp-\bk)\Delta_{\mu\nu}(\bk')F(\bk-\bk')=\frac{1}{2N}\sum_{\bk'}\Delta_{\lambda\sigma}(\bk')F(\bp-\bk') \,.
	\end{equation}
	Varying $S_2$ leads to:
	\begin{align}
	\frac{1}{\beta}\frac{\delta S_2}{\delta \Delta^*_{\sigma\lambda}(\bp)}&=\frac{1}{2\beta N}\sum_{\bk\omega_n}\mathrm{tr}\left(\frac{\delta\log(\mathcal{G}^{-1}(\bk,\i\omega_n))}{\delta\mathcal{G}^{-1}(\bk,\i\omega_n)}\frac{\delta\mathcal{G}^{-1}(\bk,\i\omega_n)}{\delta \Delta^*_{\sigma\lambda}(\bp)}\right) \nonumber\\
	&=\frac{1}{2\beta N}\sum_{\bk\omega_n}\left[\mathcal{G}(\bk,\i\omega_n)\delta^{(d)}(\bp-\bk)\begin{pmatrix} 0 & 0 \\ \delta_{\sigma\lambda} & 0 \end{pmatrix}\right] \nonumber\\
	&=\frac{1}{2\beta}\sum_{\omega_n}\left[\mathcal{G}(\bp,\i\omega_n)\right]_{\lambda\sigma}^{12} \,.
	\label{eqn:dS2}
	\end{align}
	$\left[\mathcal{G}(\bp,\i\omega_n)\right]_{\lambda\sigma}^{12}$ refers to the $(\lambda,\sigma)$ matrix element of the $(1,2)$ block (i.e., top right block) of the $\mathcal{G}$ matrix. In order to invert a matrix containing square block matrices, we make use of the following identity:
	
	\begin{equation}
	\begin{pmatrix} A & B \\ 
	C & D\end{pmatrix}^{-1}=\begin{pmatrix} A^{-1}+A^{-1}B(D-CA^{-1}B)^{-1}CA^{-1} & -A^{-1}B(D-CA^{-1}B)^{-1} \\ 
	-(D-CA^{-1}B)^{-1}CA^{-1} & (D-CA^{-1}B)^{-1}\end{pmatrix} \,.
	\end{equation}
	Focusing on the top right corner, we obtain after some algebra and using the particle-hole symmetry of the Hamiltonian:
	\begin{eqnarray}
	\left[\mathcal{G}(\bp,\i\omega_n)\right]^{12}&=&-[\omega_n^2+\i\omega_n(H_0(\bp)-\Delta(\bp) H_0^\mathrm{T}(-\bp)\Delta^{-1}(\bp))+\Delta(\bp)H_0^\mathrm{T}(-\bp)\Delta^{-1}(\bp)H_0(\bp)+\Delta(\bp)\Delta^\dagger(\bp)]^{-1}\Delta(\bp)\nonumber\\
	&=&-(\i\omega_n\mathbb{I}-\mathcal{E}(\bp))^{-1}(\i\omega_n\mathbb{I}+\mathcal{E}(\bp))^{-1}\Delta(\bp)
	\end{eqnarray}
	and $\mathcal{E}^2(\bp)\equiv H_0(\bp)H^\dagger_0(\bp)+\Delta(\bp)\Delta^\dagger(\bp)$. This matrix can be diagonalized via $\mathcal{E}^2(\bp)=U(\bp)E^2(\bp)U^\dagger(\bp)$ and the Matsubara frequency summation performed:
	\begin{align}
	\frac{1}{\beta}\frac{\delta S_2}{\delta \Delta^*_{\sigma\lambda}(\bp)}&=\frac{1}{2\beta}\sum_{\i\omega_n}-(\i\omega_n\mathbb{I}-\mathcal{E}(\bp))_{\sigma m}^{-1}(\i\omega_n\mathbb{I}+\mathcal{E}(\bp))_{m\alpha}^{-1}\Delta_{\alpha\lambda}(\bp) \nonumber\\
	&=-\frac{1}{2\beta}\sum_{\i\omega_n}U_{\sigma m}(\bp)(\i\omega_n-E(\bp))_{m}^{-1}(\i\omega_n+E(\bp))_m^{-1}U_{m\alpha}^*(\bp)\Delta_{\alpha\lambda}(\bp) \nonumber\\
	&=\frac{1}{2}U_{\sigma m}(\bp)\frac{1}{2E_m(\bp)}\tanh\left(\frac{\beta E_m(\bp)}{2}\right)U_{m\alpha}^*(\bp)\Delta_{\alpha\lambda}(\bp)
	\end{align}
	with implied sums over repeated indices. In matrix notation:
	\begin{equation}
	\frac{1}{\beta}\frac{\delta S_\mathrm{eff}}{\delta \Delta^\dagger(\bp)}=\frac{1}{2N}\sum_{\bk'} F(\bp-\bk')\Delta(\bk')+\frac{1}{2}U(\bp)\frac{1}{2E(\bp)}\tanh\left(\frac{\beta E(\bp)}{2}\right)U^\dagger(\bp)\Delta(\bp)=0 \,.
	\end{equation}
	For an inversion symmetric scattering potential $\mathcal{V}(\br)=\mathcal{V}(-\br)$, this can be rewritten in terms of the Fourier transform $V(\bk)$:
	\begin{align}
	\Delta_{\mu\nu}(\bk)&=-\frac{1}{N}\sum_{\bk'}\sum_{m\lambda} V(\bk-\bk') U_{\mu m}(\bk')\frac{1}{2E_m(\bk')}\tanh\left(\frac{E_m(\bk')}{2T}\right)U^*_{m\lambda}(\bk')\Delta_{\lambda\nu}(\bk') \nonumber\\
	&=\sum_{\bk'}V(\bk-\bk')M_{\mu\lambda}(\bk')\Delta_{\lambda\nu}(\bk') \,,
	\end{align}
	with $M_{\mu\lambda}(\bk')\equiv-\frac{1}{N}\sum_m U_{\mu m}(\bk')\frac{1}{2E_m(\bk')}\tanh\left(\frac{E_m(\bk')}{2T}\right)U^*_{m\lambda}(\bk')$. Numerically, it is useful to write this as a matrix equation at temperature $T$:
	\begin{equation}
	\vec{\Delta}(T)=\mathcal{M}(T)\;\vec{\Delta}(T)\iff\Delta_i(T)=\mathcal{M}_{ij}(T)\Delta_j(T) \,.
	\end{equation}
	For a Hamiltonian comprising $b$ bands and discretized on a momentum mesh with $N$ points, the indices ${i,j\in[1,...,b^2N]}$ which makes the matrix $\mathcal{M}$ of dimensions $b^2N\times b^2N$. More explicitly:
	\begin{equation}
	\begin{pmatrix}
	\Delta_{11}(\bk_1) \\ \Delta_{12}(\bk_1) \\ \vdots \\ \Delta_{1b}(\bk_1) \\ \vdots \\ \Delta_{bb}(\bk_1) \\ \vdots \\ \Delta_{bb}(\bk_N)
	\end{pmatrix}
	\!\!=\!\!
	\begin{pmatrix}
	V(\bk_1-\bk_1)[M(\bk_1)]\otimes\mathbb{I}_b & V(\bk_1-\bk_2)[M(\bk_2)]\otimes\mathbb{I}_b & \hdots & V(\bk_1-\bk_N)[M(\bk_N)]\otimes\mathbb{I}_b\\
	V(\bk_2-\bk_1)[M(\bk_1)]\otimes\mathbb{I}_b & V(\bk_2-\bk_2)[M(\bk_2)]\otimes\mathbb{I}_b & \hdots & V(\bk_2-\bk_N)[M(\bk_N)]\otimes\mathbb{I}_b\\
	\vdots & \vdots & \ddots & \vdots \\
	V(\bk_N-\bk_1)[M(\bk_1)]\otimes\mathbb{I}_b & V(\bk_N-\bk_2)[M(\bk_2)]\otimes\mathbb{I}_b & \hdots & V(\bk_N-\bk_N)[M(\bk_N)]\otimes\mathbb{I}_b
	\end{pmatrix}
	\!\!
	\begin{pmatrix}
	\Delta_{11}(\bk_1) \\ \Delta_{12}(\bk_1) \\ \vdots \\ \Delta_{1b}(\bk_1) \\ \vdots \\ \Delta_{bb}(\bk_1) \\ \vdots \\ \Delta_{bb}(\bk_N)
	\end{pmatrix} \,,
	\label{eqn:matrixDelta}
	\end{equation}
	where $\mathbb{I}_b$ is the $b\times b$ identity matrix. 
	
	\begin{figure*}[t] %figure moved for better placement
	\centering
	\begin{overpic}[width=\textwidth]{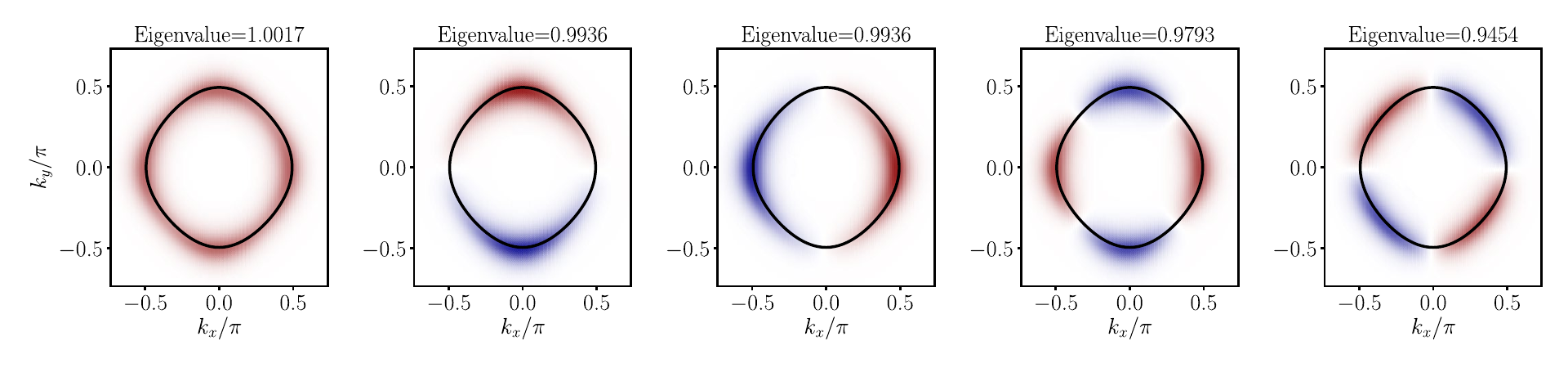}\end{overpic}
	\caption{Eigenvectors of $\mathcal{M}$ with their corresponding eigenvalues at $T=T_c\approx0.164$ for $(t_1,t_2,g_0,\ell,\bar{n})=(1,1,1,5,0.36)$. The Fermi surface is shown in black, while the colormap goes from blue (negative) to red (positive), with white being zero intensity. The largest eigenvalue corresponds to lowest energy state.}
	\label{fig:gapTc1B}
\end{figure*}

	At $T=T_c$, $E_m(\bk)\rightarrow\xi_m(\bk)$ and $U_{\mu m}(\bk)\rightarrow W_{\mu m}(\bk)$ where $\xi_m(\bk)=W^*_{m\lambda}(\bk)H_0^{\lambda\sigma}(\bk)W_{\sigma m}(\bk)$ such that $M$ no longer depends on $\Delta$. Equation \eqref{eqn:matrixDelta} reduces to an eigenvalue equation and $T_c$ is obtained when the largest eigenvalue of $\mathcal{M}$ reaches 1 (for $T\ll T_c$ all the eigenvalues are larger than 1 while for $T\gg T_c$ all the eigenvalues vanish.) 
	
	At $T=0$, the equation is nonlinear as the matrix $M$ depends on $\Delta$ and we must solve by (i) guessing an initial $\Delta^{(0)}(\bk)$, (ii) diagonalizing $H_0(\bk)H_0^\dagger(\bk)+\Delta^{(0)}(\bk)[\Delta^{(0)}(\bk)]^\dagger=U(\bk)[E^{(0)}(\bk)]^2U^{-1}(\bk')$, (iii) constructing the $\mathcal{M}$ matrix and (iv) multiplying by the `vectorized' $\vec{\Delta}^{(0)}$ to obtain a new vector $\vec{\Delta}^{(1)}$ which can then be used to repeat the procedure. The gap function $\Delta(T=0)$ is the `fixed point' of this equation.
	
	Finally, since the instability is expected near the Fermi momenta (i.e., where the denominator $E_m(\bk)\rightarrow 0$) it is useful to only store momenta within a given range of $\bk_F$. The length scale $\ell^{-1}$ provides a natural cutoff and we found that only keeping momenta within $\pm 3\ell^{-1}$ of $\bk_F$ is sufficient to reach convergent results.
\end{widetext}

%%%%%%%%%%%%%%%%%%%%%%%%%%%%%%%%%%%%%%%
% Momentum structure of the gap
%%%%%%%%%%%%%%%%%%%%%%%%%%%%%%%%%%%%%%%

\section{Eigenfunctions of the gap equation}
\label{sec:appendix:eigenfunctions}
At $T=T_c$, the spectrum of $\mathcal{M}$ tells us about the eigenmodes of the gap. The largest eigenvalue is the lowest energy state into which the system will condense first and the corresponding eigenvector shows the momentum dependence of the gap. When $\ell\rightarrow\infty$, all eigenvalues converge to 1 and the corresponding eigenvectors become localized to single momentum points on the Fermi surface. In that case, any linear combination of the eigenvectors would be a solution to the gap equation and all momenta condense simultaneously. However, for finite values of $\ell$, we generally have one eigenvalue reaching unity before the others. The subleading instabilities become closer to the leading instability at high densities (as this is equivalent to increasing $\ell$). For a given density and pairing lengthscale shown in Fig.~\ref{fig:gapTc1B}, we show that the largest eigenvalue is nondegenerate and has $s$-wave symmetry, while the next eigenvalues are doubly degenerate with $p_x$ and $p_y$ symmetry and the next two have $d$-wave symmetry. 

\section{Anisotropic dispersions and $\ell$ dependence}

\begin{figure}
	\centering
	\begin{overpic}[width=\linewidth]{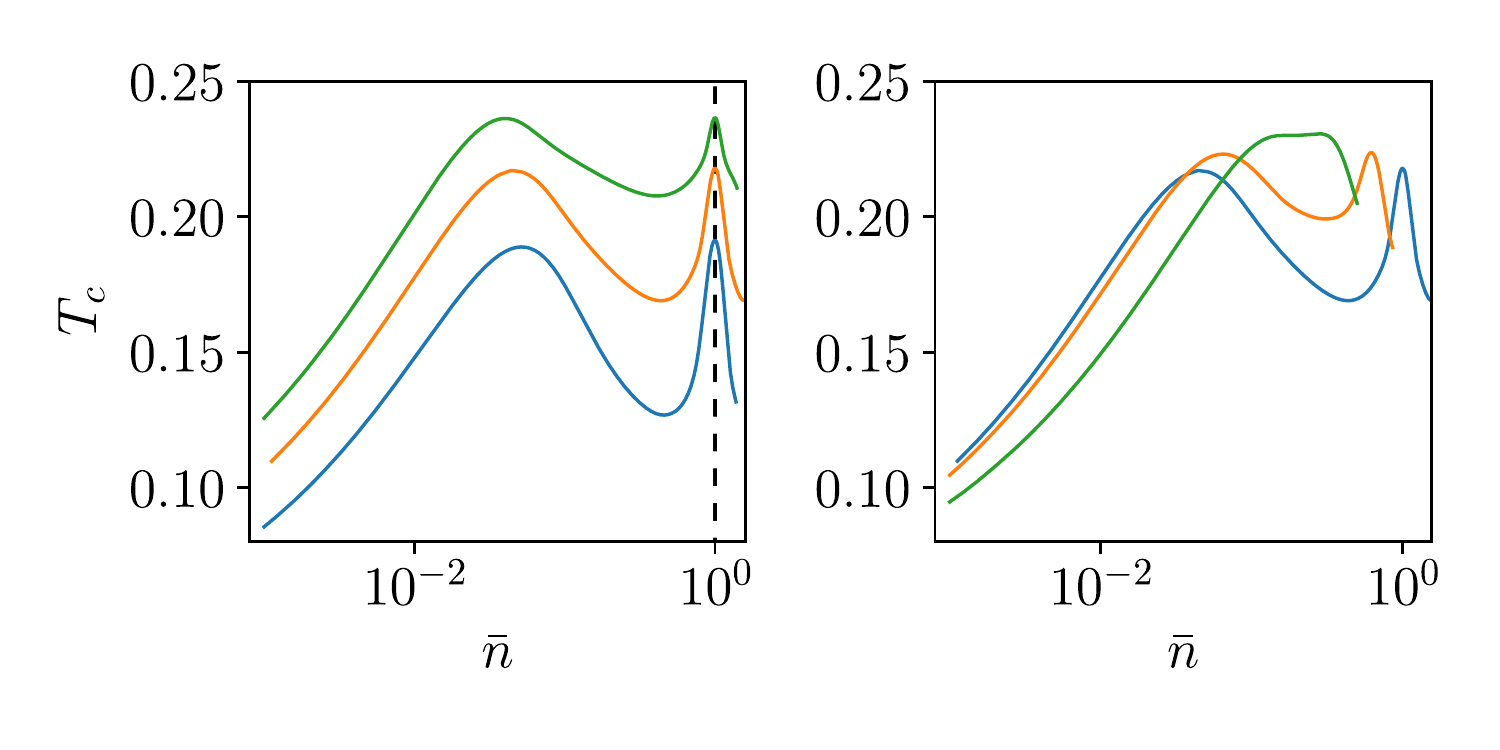}\put(16,47){(a)} \put(62,47){(b)} \end{overpic}
	\caption{Critical temperature $T_c$ as a function of electron density $\bar{n}$ for (a): $\ell=3$ (blue), $\ell=5$ (orange), and $\ell=10$ (green). The dashed vertical line at $\bar{n}=1$ shows, in contrast to the geometric peak, that the location of the van Hove peak doesn't depend on $\ell$. In (b), we observe the two peaks merge as the FS is made increasingly anisotropic, going from $t_2=t_1$ (blue) to $t_2=0.5t_1$ (orange) to $t_2=0.1t_1$ (green). }
	\label{fig:Tc1B2DLAni}
\end{figure}
As we tune the value of $\ell$ we observe that as $\ell$ is increased the location of the geometric peak moves to lower densities as expected (thus keeping $k_F^\star\ell\sim1$). On the other hand, although the van Hove peak remains at half filling [Fig. \ref{fig:Tc1B2DLAni}(a)], its tail gets sharper for small $\ell$ because Umklapp scattering only kicks in at increasingly large fillings. For an anisotropic FS with dispersion $\xi(\bk)=-2t_1\cos(k_x)-2t_2\cos(k_y)-\mu$ and $t_1\neq t_2$, the location of the van Hove peak shifts to lower densities since the elongated part of the elliptical FS hits the BZ boundary at densities $\bar{n}<1$. For a very elliptical FS ($t_2/t_1\ll 1$ or $t_2/t_1\gg1$), the van Hove point is at smaller densities and $k_F$ acquires a strong angular modulation ${k_F^{\mathrm{long}}\gg k_F^{\mathrm{short}}}$. In turn, this introduces a new condition that as soon as $k_F^{\mathrm{long}}\ell\gg1$, the available phase space for scattering starts to decrease even if ${k_F^{\mathrm{long}}\ell\lesssim1}$. In this scenario, the van Hove peak shifts to lower densities while the geometric peak is pushed to higher densities and the two eventually merge into a single peak [Fig. \ref{fig:Tc1B2DLAni}(b)]. 
%%%%%%%%%%%%%%%%%%%%%%%%%%%%%%%%%%%%%%%
% FRG flow equations
%%%%%%%%%%%%%%%%%%%%%%%%%%%%%%%%%%%%%%%

\section{FRG flow equations}
\label{sec:appendix:frg}
\begin{figure}
	\centering
	\includegraphics[width=0.9\linewidth]{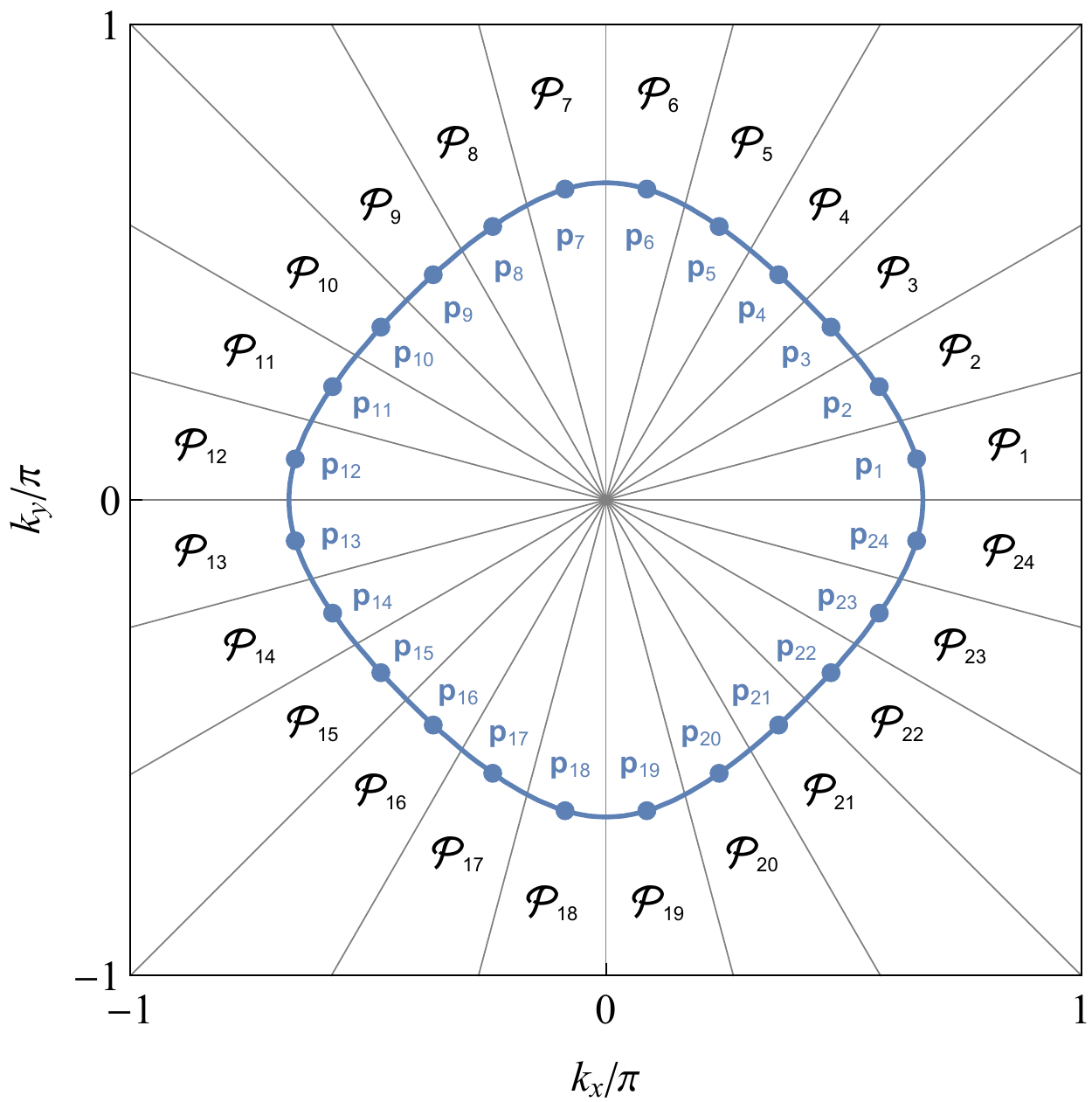}
	\caption{Partitioning of the Brillouin zone into a set of patches $\mathcal{P}_1,\dots,\mathcal{P}_N$, illustrated for $N=24$. All momentum points within a single patch $\mathcal{P}_m$ are projected onto the representative momentum $\bp_m$ which lies on the Fermi surface. }
	\label{fig:appendix:patching}
\end{figure}

The general form of the effective two-particle interaction vertex is given by 
\begin{equation}
\label{eq:appendix:frg:vertex}
V(K_1',K_2';K_1,K_2) \,,
\end{equation}
where the parameters $K_n$ are composite indices denoting tuples $(\bk_n,\omega_n,\alpha_n)$ of momentum, Matsubara frequency, and spin, respectively. 
In this most general form, the fermionic interaction vertex must be antisymmetric under the pairwise exchange of its arguments. 
For the SU(2)-symmetric model at hand, however, it is more convenient to constrain the effective interaction to a form which is inherently encoded to be SU(2) symmetric; to this end, we parametrize the effective interaction by two terms -- spin-conserving and spin-exchange terms -- which span a full basis for SU(2) invariant interactions: 
\begin{align}
V(K_1',K_2';K_1,K_2)&= U(k_1',k_2';k_1,k_2)\delta_{\alpha_1' \alpha_1}\delta_{\alpha_2'\alpha_2} \nonumber\\
&- U(k_1',k_2';k_2,k_1) \delta_{\alpha_1' \alpha_2}\delta_{\alpha_2'\alpha_1} \,.
\end{align}
Here, the composite indices $k_n$ denote pairs $(\bk_n,\omega_n)$ of momentum and Matsubara frequency, while the spin index $\alpha$ is written out explicitly. The basis function $U(k_1',k_2';k_1,k_2)$ is symmetric under simultaneous exchange of ingoing and outgoing indices. 

For further simplification of the vertex parametrization we resort to the momentum space patching approximation outlined in Ref.~\onlinecite{Halboth2000}, which is suitable in the weak coupling limit. 
In this approximation, the frequency dependence of the vertex is neglected, while the momentum dependence is parametrized such that it resolves the angular dependence around the Fermi surface, but it neglects any dependence in the radial direction. 
This is achieved by partitioning the Brillouin zone into a set of \emph{patches} $\{\mathcal{P}_1 \dots \mathcal{P}_N\}$ as shown in Fig.~\ref{fig:appendix:patching}, and projecting all momentum points within a patch $\mathcal{P}_m$ onto a single representative point $\bp_m$ on the Fermi surface, i.e., the vertex function is assumed to be constant within the entire patch. 
The parametrization of the vertex function can thus be written as
\begin{align}
&U(k_1',k_2';k_1,k_2) = \sum\limits_{i_1, i_2, i_3} u(n_{i_1}, n_{i_2}, n_{i_3}) \nonumber\\
&\quad\times \delta(\bk_1'+\bk_2'-\bk_1-\bk_2) \delta_{\bk_1' \in \mathcal{P}_{i_1}} \delta_{\bk_2' \in \mathcal{P}_{i_2}} \delta_{\bk_1 \in \mathcal{P}_{i_3}} \,,
\end{align}
where the indices $n_m$ enumerate momentum patches and the symbol $\delta_{q\in \mathcal{P}_n}=1$ if momentum $q$ lies within patch $\mathcal{P}_n$ and zero otherwise. 

\begin{figure}
	\centering
	\includegraphics[width=\linewidth]{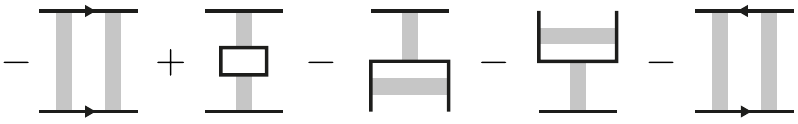}
	\caption{Diagrammatic representation of contributions to the flow of the effective interaction. Singling out particle-particle scattering (first term) is equivalent to a Bethe-Salpeter like resummation and generates results on the level of self-consistent mean-field theory. The inclusion of additional direct particle-hole (terms two to four) and crossed particle-particle (last term) scattering allows to model the interplay of competing interaction channels.}
	\label{fig:appendix:FRGchannels}
\end{figure}

The FRG flow equations are obtained by introducing an additional dependence of the interaction vertex on some RG cutoff. 
We follow the temperature flow RG scheme outlined in Ref.~\cite{Honerkamp2001}, where the temperature itself assumes to role of the RG cutoff and the flow equations take the form
\begin{equation}
\frac{\mathrm{d}}{\mathrm{d}T}u_T(n_1, n_2, n_3) = \mathcal{T}_{PP,T} + \mathcal{T}_{PH,T}^d + \mathcal{T}_{PH,T}^c \,,
\end{equation}
where the three interaction channels (particle-particle, direct particle-hole, and crossed particle-hole interaction, respectively) are given by (terms in the same order as shown in Fig.~\ref{fig:appendix:FRGchannels})

\begin{align}
&\mathcal{T}_{PP,T}(n_1, n_2, n_3) = -\sum\limits_{n} L^+_T(n, n_1+n_2) \nonumber\\
&\quad\times u_T(n_1, n_2, n) u_T(n, -n+n_1+n_2, n_3)  \nonumber\\
&\mathcal{T}^d_{PH,T}(n_1, n_2, n_3) = -\sum\limits_{n} L^-_T(n, n_1-n_3) \nonumber\\
&\quad\times \Big( -2u_T(n_1,n,n_3) u_T(n+n_1-n_3,n_2,n) \nonumber\\ 
&\quad\quad +u_T(n_1,n,n+n_1-n_3) u_T(n+n_1-n_3,n_2,n) \nonumber\\ 
&\quad\quad +u_T(n_1,n,n_3) u_T(n_2,n+n_1-n_3,n) \Big)  \nonumber\\
&\mathcal{T}^c_{PH,T}(n_1, n_2, n_3) = -\sum\limits_{n} L^-_T(n, n_2-n_3) \nonumber\\
&\quad\times u_T(n_1,n+n_2-n_3,n) u_T(n,n_2,n_3)  \,.
	\end{align}

The internal propagator bubble is defined as 
\begin{equation}
L^\pm_T(n,m) = \int_{\bk\in\mathcal{P}_n} \mp \frac{\lambda\left(\xi(\bk)\right)\pm \lambda\left(\xi(\mp \bk + \bp_m)\right)}{\xi(\bk)\pm\xi(\mp \bk + \bp_m)} \,,
\end{equation}
where $\xi(\bk)$ is the dispersion of the noninteracting system and $\lambda(\xi)$ is the temperature derivative of the Fermi distribution function
$\lambda(\xi) = \xi \e^{\xi/T}[T^2 \left( \e^{\xi/T}+1 \right)^2]^{-1}$.

In this form, the flow equations can be solved numerically to connect the high-temperature limit, in which the effective interaction vertex Eq.~\eqref{eq:appendix:frg:vertex} equals the bare interaction as defined by the Hamiltonian $\mathcal{H}_\mathrm{int}$, to the effective low-energy theory.

%%%%%%%%%%%%%%%%%%%%%%%%%%%%%%%%%%%%%%%
% Hubbard model
%%%%%%%%%%%%%%%%%%%%%%%%%%%%%%%%%%%%%%%

\section{Hubbard model}
\label{sec:appendix:hubbard}
\begin{figure}
	\centering
	\includegraphics[width=\linewidth]{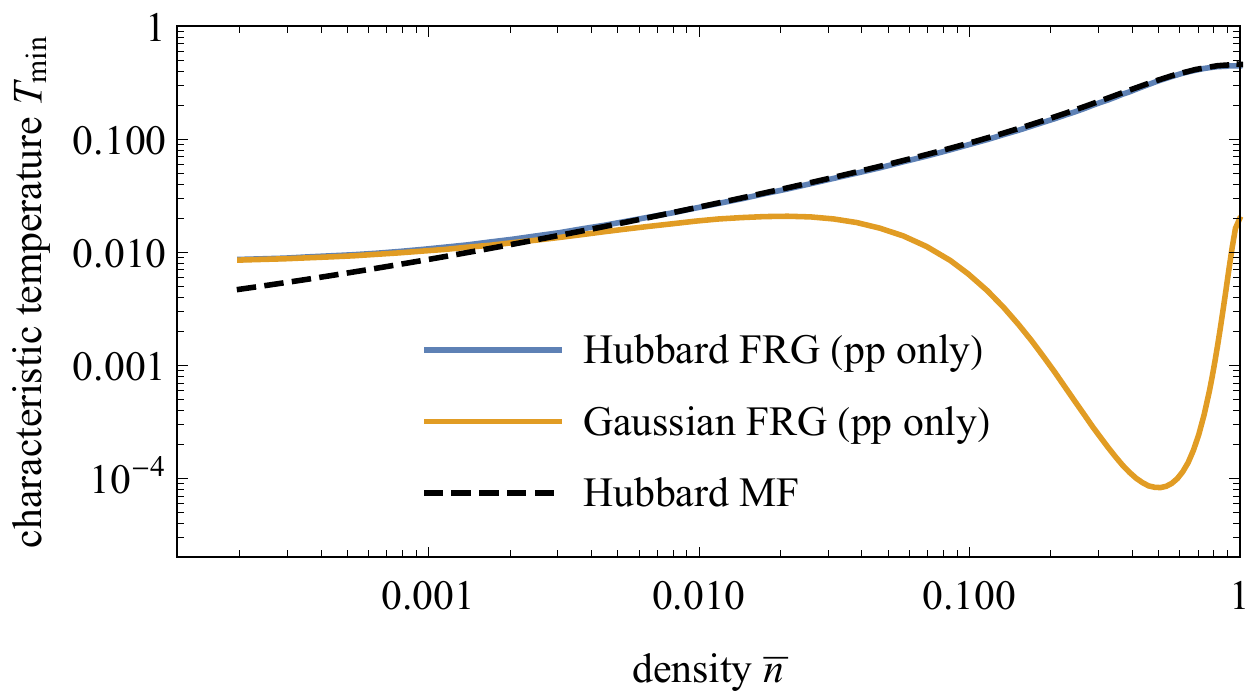}
	\caption{Benchmark of the characteristic temperature $T_\mathrm{min}$, as determined in FRG calculations (including only particle-particle scattering channel), with the mean-field critical temperature of the Hubbard model. The Hubbard limit $\ell=0$ shows excellent agreement for fillings $\bar{n}\gtrsim 10^{-3}$. The generalized Gaussian potential ($\ell=1$) approaches the Hubbard limit in the very dilute regime. }
	\label{fig:appendix:hubbard:hubbardFRG}
\end{figure}
The FRG flow equations derived in Sec.~\ref{sec:appendix:frg} are suited for the weak-coupling limit~\cite{Honerkamp2001}. 
In the dilute limit, however, when the Fermi energy scale becomes small compared to the interaction potential, the weak-coupling scenario may be violated. 
In order to convince ourselves that the approach produces meaningful results nevertheless, we benchmark the implementation against the mean-field solution of the Hubbard model at small densities. 

To this end, we consider again the general Gaussian potential introduced in the main article and set $\ell = 0$, while fixing the prefactor to $g_0=\frac{3}{2\pi}$. 
As displayed in Fig.~\ref{fig:appendix:hubbard:hubbardFRG}, the characteristic temperature scale $T_\mathrm{min}$ obtained from the FRG solution is in excellent agreement with the critical temperature as determined by the mean-field approach. Only at extremely low densities, below fillings relevant for our studies of geometric domes of $T_c$, deviations manifest. 
The FRG results remain consistent when a finite $\ell=1$ is considered in the sense that the result smoothly connects to the Hubbard limit in the dilute limit where $k_F \ell \ll 1$. This is to be expected since the width of the Gaussian profile becomes large compared to the size of the Fermi surface and the interaction potential effectively appears almost constant.

\end{document}